\documentclass[reprint, preprintnumbers, aps, prl, floatfix, superscriptaddress, nofootinbib]{revtex4-2}

\usepackage{silence}
\usepackage{amssymb, amsmath, physics, graphicx, xcolor, orcidlink, fontawesome, hyperref, bm, mathrsfs}

\newcommand{\DAMTP}{Department of Applied Mathematics and Theoretical Physics, Centre for Mathematical Sciences, University of Cambridge, Wilberforce Road, Cambridge, CB3 0WA, UK}
\newcommand{\SJC}{St John's College, Cambridge, CB2 1TP, United Kingdom}

\begin{document}

\title{Searching for Massive Scalar Fields with non-minimal Curvature Couplings in LIGO–Virgo–KAGRA Ringdown Signals}

\author{Adrian Ka-Wai Chung\,\orcidlink{0000-0003-2020-3254}}
\email{kwc43@cam.ac.uk}
\affiliation{\SJC} \affiliation{\DAMTP}

\date{\today}

\begin{abstract}
We search for massive scalar fields with non-minimal curvature couplings using the loudest ringdown signals observed by LIGO, Virgo, and KAGRA. 
We find that current ringdown observations do not provide informative constraints on the scalar-field mass. 
We nevertheless constrain the coupling length scale for scalar fields satisfying $M\mu<1$, where $\mu$ is the scalar-field mass, yielding marginalized $90\%$ credible upper limits of 60.4 km and 59.8 km for dynamical Chern--Simons and scalar Gauss--Bonnet couplings, respectively.
\end{abstract}

\maketitle

{\textit{\textbf{Introduction}}---}
The particle nature of dark matter remains unknown despite extensive astrophysical and cosmological evidence for its gravitational effects \cite{Sofue:2000jx, Bertone:2016nfn}.
Among the possible candidates are light scalar fields~\cite{Weinberg_1977, Wilczek_1978, Hui_2017, Ferreira_2021}, whose Compton wavelengths can become comparable to the characteristic length scales of astrophysical black holes.
Scalar fields in this mass range can give rise to a variety of phenomena around black holes, most notably the formation of scalar-field configurations through superradiance~\cite{Starobinskii:1973vzb, Brito_2020, Alexander_2022, Detweiler_1980, Dolan_2007, Siemonsen_2020, Baryakhtar_2021, Ng_2021, Tong_2022, Fan_2024, May_2025}.
Black-hole systems therefore provide a way to search for such particles through gravitational phenomena, including the gravitational waves produced during compact-binary coalescences~\cite{Arvanitaki_2011, Arvanitaki_2015, Berti:2019wnn, Baryakhtar:2017ngi, Brito:2017wnc, Brito:2017zvb, DAntonio:2018sff, Ghosh:2018gaw, Xie_2025}, extreme-mass-ratio inspirals~\cite{Maselli_2021, Duque_2024, Wade_2026, Rahman_2026, Lui_2026c, Li:2026foy}, and stochastic gravitational-wave backgrounds~\cite{Tsukada:2018mbp, Tsukada:2020lgt}.
Furthermore, the strong curvature sourced by black holes enables interactions to be probed that may be difficult to access in weak-field environments, complementing searches for dark matter in particle-physics experiments.

In this work, we consider massive scalar fields with non-minimal dynamical Chern--Simons and scalar Gauss--Bonnet couplings. Such curvature couplings arise in grand-unification and quantum-gravity frameworks~\cite{Cano:2019ore, Cano:2021rey, Alexander_2009, Jackiw:2003pm}. 
They allow massive scalar fields to sustain stationary and axisymmetric configurations around black holes without relying on superradiance. 
These configurations are regular at the event horizon and decay rapidly in the far field. 
Together with the curvature couplings, the scalar-field profiles modify both the black-hole spacetime and its spectrum of gravitational perturbations, while the finite scalar mass introduces an additional scale through the dimensionless combination $M\mu$. 
Recent developments in black-hole spacetime constructions \cite{Lam_2026, Lam:2025fzi, Chung_2026b, Li:2026hpw} and perturbation theory~\cite{Chung_2023, Chung_2024b, Chung:2024vaf, Chung:2025gyg, Li:2022pcy, Li:2023ulk, Li:2025fci, Weller:2026jqu, Blazquez-Salcedo:2023hwg, Blazquez-Salcedo:2024oek, Khoo:2024agm, Khoo:2024yeh, Cano:2023jbk} have enabled the computation of the gravitational quasinormal-mode frequencies of rotating black holes surrounded by such massive scalar configurations. 
The corresponding spectra have recently been obtained in the regime $M\mu<1$~\cite{Lui:2026osx}. 
They depend on both the curvature-coupling strength and the scalar-field mass, making it possible to search directly for these massive scalar fields using observed black-hole ringdowns.

Here we use the recently computed quasinormal-mode spectra of rotating black holes surrounded by massive scalar fields with these curvature couplings~\cite{Lui:2026osx} to perform such a search using some of the loudest ringdown signals observed by the LIGO--Virgo--KAGRA detectors.
We jointly infer the coupling length scale $\ell$ and scalar-field mass $\mu$, while accounting for the parameter-dependent validity domain of the perturbative quasinormal-mode model.
We find that current ringdown observations do not provide informative constraints on $\mu$.
They nevertheless constrain the coupling length scale to approximately $60\,{\rm km}$ for scalar fields satisfying $M\mu<1$ for both couplings.
Our results provide a direct observational test of massive curvature-coupled scalar fields using gravitational-wave ringdowns and quantify which properties of these dark-matter candidates can already be probed with current observations.

Throughout this work, we use geometrized units in which $c=G=\hbar=1$.

{\textit{\textbf{Massive scalar-field models}}---}We consider two massive spin-0 fields coupled to quadratic curvature invariants, following Refs.~\cite{Lui:2026osx, Cano:2019ore}. 
The theory is described by the Lagrangian density
\begin{equation}
\label{eq:Lagrangian}
\begin{split}
16\pi \mathscr{L}
={}R & +\ell_1{}^2 \vartheta_1 \mathscr{G}
-\frac{1}{2}\nabla_\nu\vartheta_1\nabla^\nu\vartheta_1
-\frac{1}{2}\mu_1{}^2\vartheta_1{}^2
\\
&+\ell_2{}^2 \vartheta_2 \mathscr{P}
-\frac{1}{2}\nabla_\nu\vartheta_2\nabla^\nu\vartheta_2
-\frac{1}{2}\mu_2{}^2\vartheta_2{}^2 .
\end{split}
\end{equation}
Here, $\vartheta_1$ is a scalar field of mass $\mu_1$ coupled to the Gauss--Bonnet invariant ($\mathscr{G}$), while $\vartheta_2$ is a pseudoscalar field of mass $\mu_2$ coupled to the Pontryagin invariant ($\mathscr{P}$). 
The corresponding coupling length scales are denoted by $\ell_1$ and $\ell_2$.
The curvature invariants in Eq.~\eqref{eq:Lagrangian} are 
\begin{equation}
\begin{split}
\mathscr{G}
&=
R^2
-4R_{\mu\nu}R^{\mu\nu}
+R_{\mu\nu\rho\sigma}R^{\mu\nu\rho\sigma}, \\
\mathscr{P}
&=
\tilde{R}^{\mu\nu\rho\sigma}R_{\mu\nu\rho\sigma},
\end{split}
\end{equation}
where $R_{\mu\nu}$ and $R_{\mu\nu\rho\sigma}$ are the Ricci and Riemann tensors, respectively, and $\tilde{R}_{\mu\nu\rho\sigma}$ denotes the dual Riemann tensor.
Setting $(\ell_1,\ell_2)=(\ell,0)$ gives the massive scalar--Gauss--Bonnet theory considered here, whereas $(\ell_1,\ell_2)=(0,\ell)$ gives massive dynamical Chern--Simons gravity.

The curvature couplings modify both the stationary black-hole spacetime and its quasinormal-mode spectrum \cite{Lui:2026osx}. 
In the small-coupling regime, the leading correction is controlled by
\begin{equation}
\zeta =
\left(\frac{\ell}{M}\right)^4
=
\left[\frac{\ell(1+z)}{M_z}\right]^4 ,
\label{eq:zeta}
\end{equation}
where $M$ and $M_z=(1+z)M$ are the source-frame and redshifted remnant masses, respectively.
For either coupling, the complex frequency of a given quasinormal mode and parity $(\rm q)$ can therefore be written as
\begin{equation}
\omega^{(\mathrm{q})}
=
\frac{1}{M}
\left[
\hat{\omega}^{(0)}(a)
+
\zeta\,
\Omega^{(\mathrm{q})}
\left(
a,
M \mu
\right)
\right],
\label{eq:QNMFs}
\end{equation}
where $\hat{\omega}^{(0)}(a)$ is the dimensionless Kerr quasinormal-mode frequency and $\Omega^{(\mathrm{q})}(a,M \mu)$ denotes the leading parity $(\rm q)$-dependent correction associated with the corresponding curvature coupling.
The finite scalar mass therefore introduces an additional dependence of the ringdown spectrum through the dimensionless combination $M \mu$.
We use the massive quasinormal-mode spectra computed in~\cite{Lui:2026osx}, where the frequency shifts and their dependence on black-hole spin, scalar mass, and perturbation parity were calculated and numerically validated.
In the limit $\mu\rightarrow0$, Eq.~\eqref{eq:QNMFs} reduces to the massless model employed in~\cite{Chung:2025ucb}.

{\textit{\textbf{Waveform model}}---}We model the ringdown signal using the parity-resolved time-domain waveform implemented in \texttt{pyRing} \cite{pyRing, Carullo:2019flw}, following~\cite{Chung:2025ucb,Maenaut:2024oci}.
The waveform construction, mode amplitudes, and treatment of the positive- and negative-frequency branches are unchanged from~\cite{Chung:2025ucb}.
The only modification relevant to the present analysis is that the massless quasinormal-mode corrections are replaced by the massive spectra described above, so that the mode frequencies depend jointly on the coupling scale $\ell$ and the scalar mass $\mu$ through $\Omega^{(\mathrm{q})}(a,M \mu)$.
As in~\cite{Chung:2025ucb}, we impose the reflection-symmetry relation between the positive- and negative-frequency amplitudes and neglect scalar-led quasinormal modes, restricting the analysis to the leading corrections to the gravitational-led spectrum.

{\textit{\textbf{Data analysis}}---}The parameter-estimation analysis closely follows the prescription of~\cite{Chung:2025ucb}, including the event selection, ringdown start time, treatment of extrinsic parameters, priors on the remnant and mode parameters, and the time-domain Bayesian analysis with \texttt{pyRing}.
We analyze GW150914, GW190521\_074359, GW200129\_065458, GW231226\_101520, and GW250114\_082203.
Unless stated otherwise below, we adopt the same analysis choices as in~\cite{Chung:2025ucb}.

To restrict the analysis to the parameter space over which the quasinormal-mode frequencies of Ref.~\cite{Lui:2026osx} are valid, we impose the following theory-validity conditions during sampling:
\begin{equation}
    \ell < M,
    \qquad
    M \mu < 1,
    \quad
    \text{and}
    \quad
    \mathrm{Im}\!\left(\omega^{(\mathrm{P})}_{nlm}\right)<0 ,
\label{eq:guard}
\end{equation}
for the relevant parity sectors and modes.
These conditions define a validity guard and, consequently, modify the nuisance-parameter volume available at fixed $(\ell,\mu)$.
The treatment of this effect in Bayesian inference and posterior sampling is described below.

For GW150914, GW190521\_074359, GW200129\_065458, and GW231226\_101520, we model the signal with the fundamental $nlm=022$ mode.
For GW250114\_082203, we additionally include the 122 overtone in the inference, motivated by the evidence for an overtone contribution reported for this event \cite{LIGOScientific:2025wao}. 
For both the 022 fundamental mode and the 122 overtone, we include the corresponding massive-scalar corrections to the axial- and polar-led quasinormal-mode frequencies, while independently inferring their complex amplitudes.

{\textit{\textbf{Bayesian Framework}}---}We infer jointly the coupling length scale $\ell$ and scalar mass $\mu$, which are common parameters across all events.
For each event $i$, we denote these parameters collectively by $\lambda=(\ell,\mu)$ and the event-specific parameters by $\boldsymbol{\eta}_i$.
The analysis follows the Bayesian prescription of~\cite{Chung:2025ucb}, with an additional treatment required by the parameter-dependent validity conditions of the massive-scalar quasinormal-mode model.

\begin{figure*}[htp!]
    \centering
    \includegraphics[width=\textwidth]
    {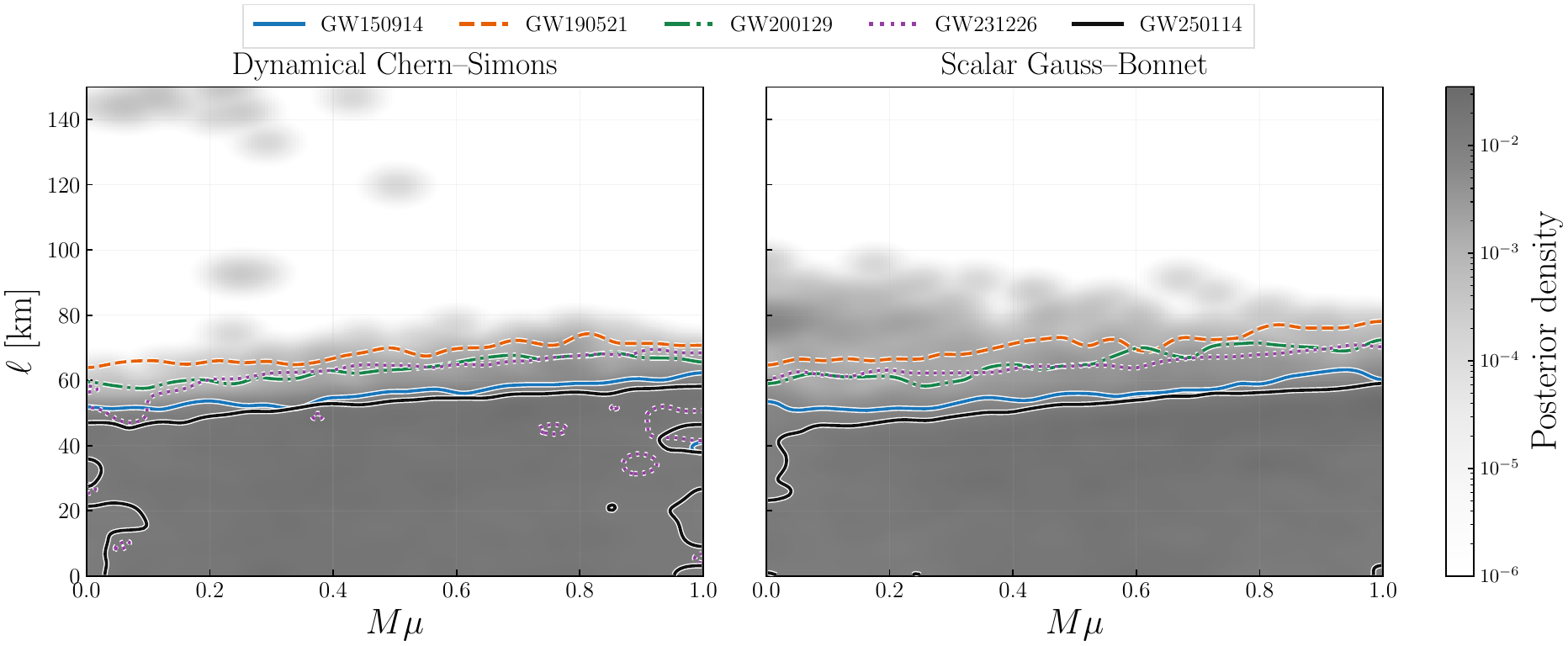}
    \caption{
    Posterior distributions in the $(\ell,M\mu)$ plane for massive scalar fields with dynamical Chern--Simons (left) and scalar Gauss--Bonnet (right) couplings, obtained from Bayesian inference of the GW250114 ringdown signal. 
    The grayscale shows the GW250114 posterior density, while the contours show the $90\%$ credible regions obtained from the individual ringdown events analyzed in this work.
    }
    \label{fig:Mmu_posteriors}
\end{figure*}

The production likelihood is evaluated only within the validity region selected by a guard $G_i(\lambda,\boldsymbol{\eta}_i)$.
At fixed $(\ell,\mu)$, the fraction of the nuisance prior that satisfies the guard is
\begin{equation}
N_i(\ell,\mu)
=
\int d\boldsymbol{\eta}_i\,
\pi_i(\boldsymbol{\eta}_i|\ell,\mu)\,
G_i(\ell,\mu,\boldsymbol{\eta}_i).
\label{eq:guard_fraction}
\end{equation}
Because $N_i$ varies across the $(\ell,\mu)$ plane, the posterior returned directly by the sampler contains a parameter-dependent prior-volume effect.
We remove this contribution by reweighting each event-level posterior according to
\begin{equation}
\hat{p}_i(\ell,\mu|d_i)
\propto
\frac{
\tilde{p}_i(\ell,\mu|d_i)
}{
N_i(\ell,\mu)
}.
\label{eq:guard_reweighting}
\end{equation}
Here $\hat{p}_i(\ell,\mu|d_i)$ is the reweighted posterior, and $\tilde{p}_i$ is the posterior constructed via nested sampling. 

Since $\ell$ and $\mu$ characterize the underlying theory, we assume that they take the same values for all observed black-hole remnants.
For statistically independent events, the joint posterior is therefore
\begin{equation}
p(\ell,\mu|\{d_i\})
\propto
\pi(\ell,\mu)
\prod_i
\hat{{\cal L}}_i(\ell,\mu),
\label{eq:combined_posterior}
\end{equation}
where $\hat{{\cal L}}_i$ denotes the event-level likelihood after conditioning on the validity region.
Equivalently,
\begin{equation}
p(\ell,\mu|\{d_i\})
\propto
\pi(\ell,\mu)^{1-N_{\rm ev}}
\prod_i
\hat{p}_i(\ell,\mu|d_i).
\label{eq:combined_posterior_from_posteriors}
\end{equation}
In this work, we adopt a common uniform prior on $(\ell,\mu)$ over the ranges $0\leq\ell\leq300\,{\rm km}$ and $0\leq\mu\leq10^{-8}\,{\rm eV}$. 
The prior factor in Eq.~\eqref{eq:combined_posterior_from_posteriors} is therefore constant within its support, so the combined posterior is proportional to the product of the reweighted event-level posteriors.

We report the $90\%$ highest-posterior-density credible region $\mathcal{C}_{90}$ of the combined two-dimensional posterior, defined by
\begin{equation}
\mathcal{C}_{90}
=
\left\{
(\ell,\mu):
p(\ell,\mu|\{d_i\})\geq p_{\star}
\right\},
\end{equation}
where the density threshold $p_{\star}$ is chosen such that
\begin{equation}
\int_{\mathcal{C}_{90}}
d\ell\,d\mu\,
p(\ell,\mu|\{d_i\})
=
0.9.
\label{eq:90_credible_region}
\end{equation}

{\textit{\textbf{Results}}---}Figure~\ref{fig:Mmu_posteriors} shows the two-dimensional posterior distributions of $\ell$ and $M\mu$ for massive scalar fields with dynamical Chern--Simons (left) and scalar Gauss--Bonnet (right) couplings, obtained from Bayesian inference of the ringdown signal of GW250114\_082203, the loudest gravitational-wave event announced to date by the LIGO--Virgo--KAGRA Collaboration~\cite{LIGOScientific:2025wao}. 
The grayscale shows the GW250114 posterior density, while the contours show the $90\%$ credible regions obtained from the individual ringdown events analyzed in this work. 
The posterior support decreases rapidly toward large $\ell$, with little support remaining by $\ell\sim80\,{\rm km}$. The marginalized posteriors of $\zeta=(\ell/M)^4$, shown in Fig.~\ref{fig:zeta_posteriors} of the Supplemental Material, decrease well before the validity boundary $\zeta=1$, indicating that the constraints on $\ell$ are not driven by truncation at the validity boundary. 
In contrast, the posterior retains substantial support up to the validity boundary $M\mu=1$, indicating that our analyses do not provide informative constraints on the scalar-field mass. 
This difference can be understood from the quasinormal-mode correction $\zeta\,\Omega^{(\mathrm{q})}(a,M\mu)$: the coupling dependence inherits the quartic scaling $\zeta\propto\ell^4$, whereas $\Omega^{(\mathrm{q})}$ varies more mildly with $M\mu$~\cite{Lui:2026osx}.

The upward trend of the $90\%$ credible contours with increasing $M\mu$ can be understood qualitatively from the parity dependence of the quasinormal-mode shifts. 
In our waveform model, the amplitudes of the axial- and polar-led branches are independent parameters \cite{Chung:2025ucb}. 
Therefore, if one branch develops a comparatively large deviation from the GR spectrum, its contribution can be suppressed through the corresponding amplitude, allowing the likelihood to remain supported by the branch whose quasinormal-mode frequency stays closer to its GR value. 
This suggests that the constraint on $\ell$ may be influenced primarily by the less strongly shifted parity branch. 
As shown in Ref.~\cite{Lui:2026osx}, the leading quasinormal-mode frequency corrections for the less strongly shifted parity branches, namely the polar-led branch in dynamical Chern--Simons gravity and the axial-led branch in scalar Gauss--Bonnet gravity, become less sensitive to changes in $M\mu$ as $M\mu$ approaches unity. 
This behavior suggests that the predicted quasinormal-mode frequencies become progressively less sensitive to variations in the scalar-field mass in this regime.

\begin{figure}[htp!]
    \centering
    \includegraphics[width=\columnwidth]
    {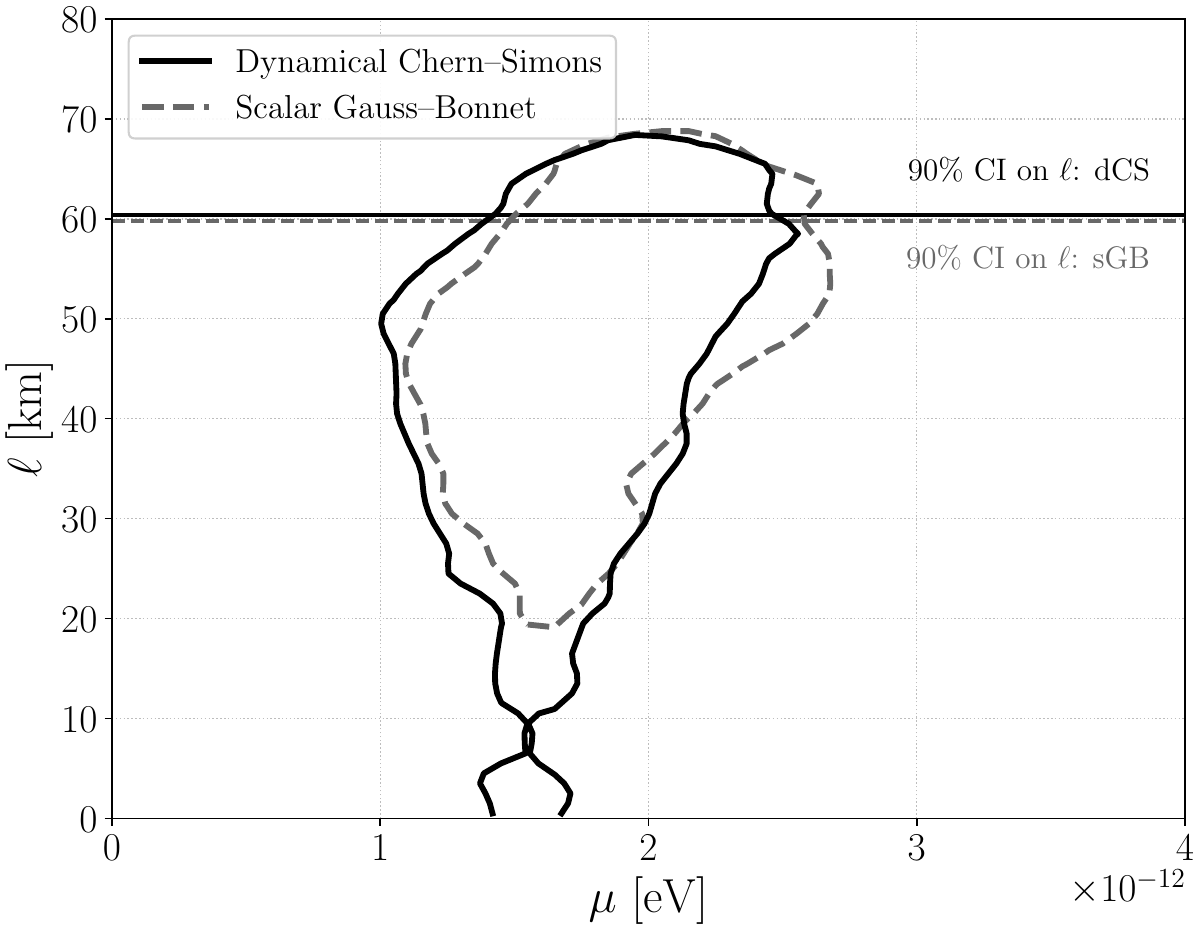}
    \caption{
    Combined $90\%$ credible regions in the $(\ell,\mu)$ plane for massive scalar fields with dynamical Chern--Simons (solid) and scalar Gauss--Bonnet (dashed) couplings. The horizontal lines mark the corresponding marginalized $90\%$ credible upper limits on the coupling length scale $\ell$.
    }
    \label{fig:combined_constraints}
\end{figure}

\begin{table}[t]
\centering
\caption{
Marginalized $90\%$ upper credible limits on the coupling length scale $\ell$ (in km) for the individual ringdown events and their combination.
``Original'' denotes the posterior obtained directly through sampling [$\tilde{p}$ in Eq.~(\ref{eq:guard_reweighting})], while ``Reweighted'' denotes the posterior corrected for the parameter-dependent validity fraction [$\hat{p}$ in Eq.~(\ref{eq:guard_reweighting})].
}
\label{tab:ell90_event_limits}
\begin{tabular}{lcccc}
\hline\hline
& \multicolumn{2}{c}{dCS} & \multicolumn{2}{c}{sGB} \\
Event
& Original
& Reweighted
& Original
& Reweighted \\
\hline
GW150914 & 56.1 & 64.6 & 55.8 & 64.4 \\
GW190521 & 71.1 & 77.9 & 73.2 & 76.5 \\
GW200129 & 63.6 & 73.0 & 66.0 & 71.1 \\
GW231226 & 61.9 & 67.3 & 65.1 & 70.5 \\
GW250114 & 51.2 & 57.5 & 52.3 & 58.8 \\\hline
Combined & 48.5 & 60.4 & 47.1 & 59.8 \\
\hline\hline
\end{tabular}
\end{table}

Table~\ref{tab:ell90_event_limits} lists the marginalized $90\%$ credible upper limits on the coupling length scale $\ell$ for massive scalar fields with dynamical Chern--Simons and scalar Gauss--Bonnet couplings, obtained from the individual ringdown signals and from their combination. 
The combined upper limits are approximately $50\%$ larger than those reported in Ref.~\cite{Chung:2025ucb} for massless scalar fields with the corresponding curvature couplings. 
This is expected because the present analysis simultaneously infers $\ell$ and the scalar-field mass $\mu$. Marginalizing over the additional parameter $\mu$ broadens the allowed range of $\ell$, particularly because the quasinormal-mode corrections depend jointly on $\ell$ and $M\mu$.

Figure~\ref{fig:combined_constraints} shows the $90\%$ credible contours of the reweighted posteriors in the $(\ell,\mu)$ plane for massive scalar fields with dynamical Chern--Simons (solid) and scalar Gauss--Bonnet (dashed) couplings, combined across the analyzed ringdown signals according to Eq.~\ref{eq:combined_posterior_from_posteriors}.
The horizontal lines mark the corresponding marginalized $90\%$ credible upper limits on the coupling length scale, $\ell_{90}=60.4\,{\rm km}$ for dynamical Chern--Simons gravity and $\ell_{90}=59.8\,{\rm km}$ for scalar Gauss--Bonnet gravity. 
Although the combined $90\%$ credible regions are confined to $\mu\lesssim3\times10^{-12}\,{\rm eV}$, this apparent upper boundary should not be interpreted as an informative constraint on the scalar-field mass. 
Rather, it primarily reflects the restrictions imposed by the validity conditions of the quasinormal-mode model, as illustrated by the broad posterior support extending toward $M\mu=1$ in Fig.~\ref{fig:Mmu_posteriors}.

\begin{figure}[htp!]
    \centering
    \includegraphics[width=\columnwidth]
    {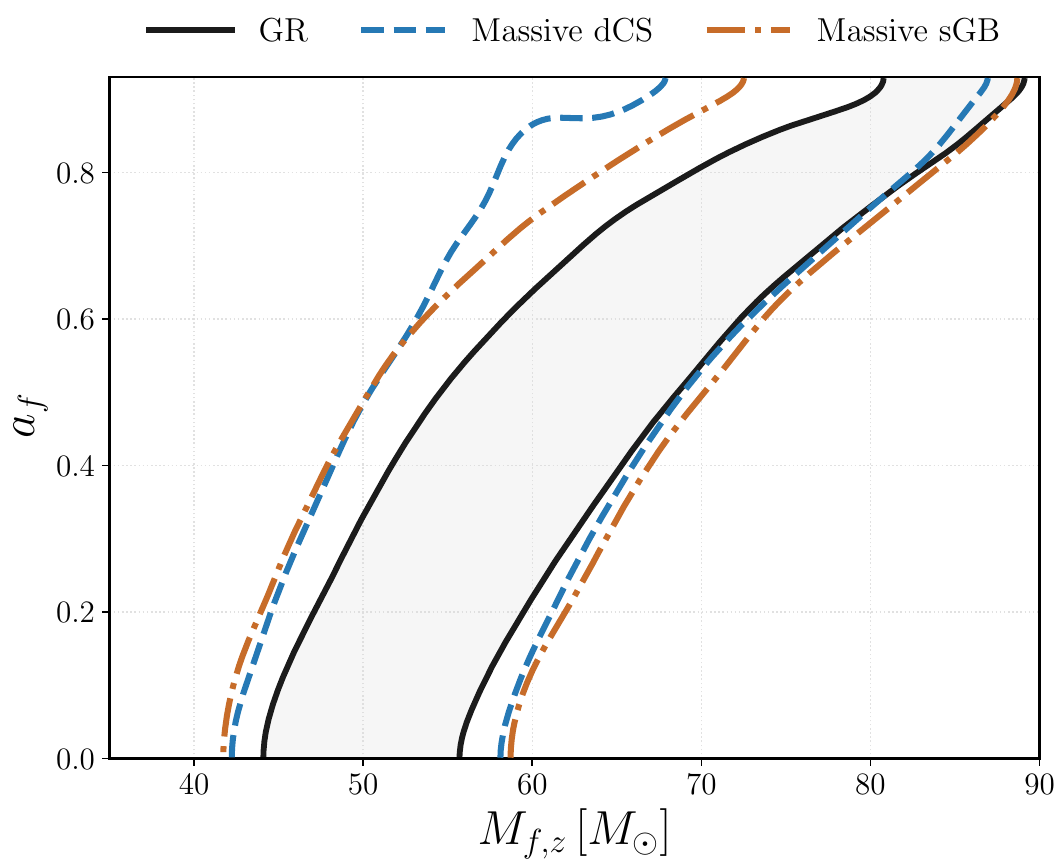}
    \caption{Marginalized $90\%$ credible regions in the detector-frame remnant mass and spin $(M_{z},a_f)$ plane, obtained from Bayesian inference of the GW150914 ringdown signal under general relativity (black solid), massive dynamical Chern--Simons gravity (blue dashed), and massive scalar Gauss--Bonnet gravity (orange dash-dotted). 
    }
    \label{fig:GW150914_CI}
\end{figure}

We conclude this section by a sanity check. 
Figure~\ref{fig:GW150914_CI} compares the marginalized $90\%$ credible regions in the detector-frame remnant mass and spin $(M_{z},a_f)$ plane inferred from GW150914 under general relativity, massive dynamical Chern--Simons gravity, and massive scalar Gauss--Bonnet gravity. 
The credible regions exhibit substantial overlap, indicating that the inferred remnant mass and spin remain consistent across the three models. 
Although the massive-scalar models introduce two additional parameters, $\ell$ and $\mu$, the breaking of isospectrality gives rise to distinct complex quasinormal-mode frequencies for the axial and polar sectors. 
These provide, effectively, four parameters that depend on $(M_{z},a_f,\ell,\mu)$, allowing additional information to distinguish the remnant parameters from the scalar-field parameters. 
However, their independent measurability depends on the excitation amplitudes and the information available in the ringdown signal. 
The overlap of the credible regions suggests that allowing for massive-scalar corrections does not substantially alter the inferred remnant parameters for this event.

{\textit{\textbf{Concluding remarks}}---}We apply the quasinormal-mode spectra of gravitational perturbations of rotating black holes surrounded by massive scalar fields with non-minimal curvature couplings, computed in Ref.~\cite{Lui:2026osx}, to some of the loudest ringdown signals detected by the LIGO--Virgo--KAGRA detectors. 
This ringdown-only analysis complements the inspiral-only search for massive scalar fields in Ref.~\cite{Xie_2025}. 
Although our analyses do not provide informative constraints on the scalar-field mass, the results reported in this work illustrate how gravitational-wave observations can connect Lagrangian densities in theories of fundamental physics to astronomical observations.

The ringdown search also suggests several extensions. 
A joint inspiral--ringdown analysis could exploit complementary information from scalar radiation during the inspiral and from the mass-dependent quasinormal-mode spectrum of the remnant, providing comprehensive search of massive scalar fields across full coalescence signals. 
Incorporating the dependence of the axial- and polar-sector amplitudes on the coupling length scale and scalar-field mass~\cite{Wagle:2026bpu} could further improve the accuracy of the search. 
More generally, the spectral-inference framework developed here could be adapted to searches for ultralight bosonic environments around stellar-mass black holes, once sufficiently accurate quasinormal-mode spectra for such systems become available.

{\textit{\textbf{Acknowledgments}}---}The author thanks Gregorio Carullo and the participants of the ``Ringdown in Lisbon'' meeting for extremely stimulating discussions.
The author acknowledges support from a Title E Fellowship at St John's College, University of Cambridge.
The author also thanks Leif Lui for insightful discussions and feedback on the initial version of the manuscript.
The calculations and results reported in this paper were produced using the computational resources of the Illinois Campus Cluster, which is a computing resource that is operated by the Illinois Campus Cluster Program (ICCP) in conjunction with National Center for Supercomputing Applications (NCSA), and the computational resources of Scientific and High Performance Computing at DAMTP, Cambridge.
This work used the large language model (LLM) \texttt{ChatGPT 5.6 Sol Extra High} to assist with literature review, coding, running scripts, documentation, and refining parts of this text, all closely supervised by A.K.W.C.

{\textit{\textbf{Data Availability}}---}The gravitational-wave strain data analyzed in this work are publicly available through the Gravitational Wave Open Science Center (GWOSC) at \url{https://gwosc.org/}.
The Bayesian inference is performed using the time-domain inference package \texttt{pyRing}, which is publicly available at \url{https://git.ligo.org/lscsoft/pyring}.
The posterior samples produced in this work are available at~\cite{10.5281/zenodo.24137075}.

\vspace{1cm}
\bibliographystyle{apsrev4-1}
\bibliography{references}

\providecommand{\noopsort}[1]{}\providecommand{\singleletter}[1]{#1}%
\begin{thebibliography}{62}%
\makeatletter
\providecommand \@ifxundefined [1]{%
 \@ifx{#1\undefined}
}%
\providecommand \@ifnum [1]{%
 \ifnum #1\expandafter \@firstoftwo
 \else \expandafter \@secondoftwo
 \fi
}%
\providecommand \@ifx [1]{%
 \ifx #1\expandafter \@firstoftwo
 \else \expandafter \@secondoftwo
 \fi
}%
\providecommand \natexlab [1]{#1}%
\providecommand \enquote  [1]{``#1''}%
\providecommand \bibnamefont  [1]{#1}%
\providecommand \bibfnamefont [1]{#1}%
\providecommand \citenamefont [1]{#1}%
\providecommand \href@noop [0]{\@secondoftwo}%
\providecommand \href [0]{\begingroup \@sanitize@url \@href}%
\providecommand \@href[1]{\@@startlink{#1}\@@href}%
\providecommand \@@href[1]{\endgroup#1\@@endlink}%
\providecommand \@sanitize@url [0]{\catcode `\\12\catcode `\$12\catcode
  `\&12\catcode `\#12\catcode `\^12\catcode `\_12\catcode `\%12\relax}%
\providecommand \@@startlink[1]{}%
\providecommand \@@endlink[0]{}%
\providecommand \url  [0]{\begingroup\@sanitize@url \@url }%
\providecommand \@url [1]{\endgroup\@href {#1}{\urlprefix }}%
\providecommand \urlprefix  [0]{URL }%
\providecommand \Eprint [0]{\href }%
\providecommand \doibase [0]{http://dx.doi.org/}%
\providecommand \selectlanguage [0]{\@gobble}%
\providecommand \bibinfo  [0]{\@secondoftwo}%
\providecommand \bibfield  [0]{\@secondoftwo}%
\providecommand \translation [1]{[#1]}%
\providecommand \BibitemOpen [0]{}%
\providecommand \bibitemStop [0]{}%
\providecommand \bibitemNoStop [0]{.\EOS\space}%
\providecommand \EOS [0]{\spacefactor3000\relax}%
\providecommand \BibitemShut  [1]{\csname bibitem#1\endcsname}%
\let\auto@bib@innerbib\@empty
\bibitem [{\citenamefont {Sofue}\ and\ \citenamefont
  {Rubin}(2001)}]{Sofue:2000jx}%
  \BibitemOpen
  \bibfield  {author} {\bibinfo {author} {\bibfnamefont {Y.}~\bibnamefont
  {Sofue}}\ and\ \bibinfo {author} {\bibfnamefont {V.}~\bibnamefont {Rubin}},\
  }\href {\doibase 10.1146/annurev.astro.39.1.137} {\bibfield  {journal}
  {\bibinfo  {journal} {Ann. Rev. Astron. Astrophys.}\ }\textbf {\bibinfo
  {volume} {39}},\ \bibinfo {pages} {137} (\bibinfo {year} {2001})},\ \Eprint
  {http://arxiv.org/abs/astro-ph/0010594} {arXiv:astro-ph/0010594} \BibitemShut
  {NoStop}%
\bibitem [{\citenamefont {Bertone}\ and\ \citenamefont
  {Hooper}(2018)}]{Bertone:2016nfn}%
  \BibitemOpen
  \bibfield  {author} {\bibinfo {author} {\bibfnamefont {G.}~\bibnamefont
  {Bertone}}\ and\ \bibinfo {author} {\bibfnamefont {D.}~\bibnamefont
  {Hooper}},\ }\href {\doibase 10.1103/RevModPhys.90.045002} {\bibfield
  {journal} {\bibinfo  {journal} {Rev. Mod. Phys.}\ }\textbf {\bibinfo {volume}
  {90}},\ \bibinfo {pages} {045002} (\bibinfo {year} {2018})},\ \Eprint
  {http://arxiv.org/abs/1605.04909} {arXiv:1605.04909 [astro-ph.CO]}
  \BibitemShut {NoStop}%
\bibitem [{\citenamefont {Weinberg}(1978)}]{Weinberg_1977}%
  \BibitemOpen
  \bibfield  {author} {\bibinfo {author} {\bibfnamefont {S.}~\bibnamefont
  {Weinberg}},\ }\href {\doibase 10.1103/PhysRevLett.40.223} {\bibfield
  {journal} {\bibinfo  {journal} {Phys. Rev. Lett.}\ }\textbf {\bibinfo
  {volume} {40}},\ \bibinfo {pages} {223} (\bibinfo {year} {1978})}\BibitemShut
  {NoStop}%
\bibitem [{\citenamefont {Wilczek}(1978)}]{Wilczek_1978}%
  \BibitemOpen
  \bibfield  {author} {\bibinfo {author} {\bibfnamefont {F.}~\bibnamefont
  {Wilczek}},\ }\href {\doibase 10.1103/PhysRevLett.40.279} {\bibfield
  {journal} {\bibinfo  {journal} {Phys. Rev. Lett.}\ }\textbf {\bibinfo
  {volume} {40}},\ \bibinfo {pages} {279} (\bibinfo {year} {1978})}\BibitemShut
  {NoStop}%
\bibitem [{\citenamefont {Hui}\ \emph {et~al.}(2017)\citenamefont {Hui},
  \citenamefont {Ostriker}, \citenamefont {Tremaine},\ and\ \citenamefont
  {Witten}}]{Hui_2017}%
  \BibitemOpen
  \bibfield  {author} {\bibinfo {author} {\bibfnamefont {L.}~\bibnamefont
  {Hui}}, \bibinfo {author} {\bibfnamefont {J.~P.}\ \bibnamefont {Ostriker}},
  \bibinfo {author} {\bibfnamefont {S.}~\bibnamefont {Tremaine}}, \ and\
  \bibinfo {author} {\bibfnamefont {E.}~\bibnamefont {Witten}},\ }\href
  {\doibase 10.1103/physrevd.95.043541} {\bibfield  {journal} {\bibinfo
  {journal} {Physical Review D}\ }\textbf {\bibinfo {volume} {95}} (\bibinfo
  {year} {2017}),\ 10.1103/physrevd.95.043541}\BibitemShut {NoStop}%
\bibitem [{\citenamefont {Ferreira}(2021)}]{Ferreira_2021}%
  \BibitemOpen
  \bibfield  {author} {\bibinfo {author} {\bibfnamefont {E.~G.~M.}\
  \bibnamefont {Ferreira}},\ }\href {\doibase 10.1007/s00159-021-00135-6}
  {\bibfield  {journal} {\bibinfo  {journal} {The Astronomy and Astrophysics
  Review}\ }\textbf {\bibinfo {volume} {29}} (\bibinfo {year} {2021}),\
  10.1007/s00159-021-00135-6}\BibitemShut {NoStop}%
\bibitem [{\citenamefont {Starobinskii}(1973)}]{Starobinskii:1973vzb}%
  \BibitemOpen
  \bibfield  {author} {\bibinfo {author} {\bibfnamefont {A.~A.}\ \bibnamefont
  {Starobinskii}},\ }\href@noop {} {\bibfield  {journal} {\bibinfo  {journal}
  {Sov. Phys. JETP}\ }\textbf {\bibinfo {volume} {37}},\ \bibinfo {pages} {28}
  (\bibinfo {year} {1973})}\BibitemShut {NoStop}%
\bibitem [{\citenamefont {Brito}\ \emph {et~al.}(2020)\citenamefont {Brito},
  \citenamefont {Cardoso},\ and\ \citenamefont {Pani}}]{Brito_2020}%
  \BibitemOpen
  \bibfield  {author} {\bibinfo {author} {\bibfnamefont {R.}~\bibnamefont
  {Brito}}, \bibinfo {author} {\bibfnamefont {V.}~\bibnamefont {Cardoso}}, \
  and\ \bibinfo {author} {\bibfnamefont {P.}~\bibnamefont {Pani}},\ }\href
  {\doibase 10.1007/978-3-030-46622-0} {\emph {\bibinfo {title} {Superradiance:
  New Frontiers in Black Hole Physics}}}\ (\bibinfo  {publisher} {Springer
  International Publishing},\ \bibinfo {year} {2020})\BibitemShut {NoStop}%
\bibitem [{\citenamefont {Alexander}\ \emph {et~al.}(2023)\citenamefont
  {Alexander}, \citenamefont {Gabadadze}, \citenamefont {Jenks},\ and\
  \citenamefont {Yunes}}]{Alexander_2022}%
  \BibitemOpen
  \bibfield  {author} {\bibinfo {author} {\bibfnamefont {S.}~\bibnamefont
  {Alexander}}, \bibinfo {author} {\bibfnamefont {G.}~\bibnamefont
  {Gabadadze}}, \bibinfo {author} {\bibfnamefont {L.}~\bibnamefont {Jenks}}, \
  and\ \bibinfo {author} {\bibfnamefont {N.}~\bibnamefont {Yunes}},\ }\href
  {\doibase 10.1103/PhysRevD.107.084016} {\bibfield  {journal} {\bibinfo
  {journal} {Phys. Rev. D}\ }\textbf {\bibinfo {volume} {107}},\ \bibinfo
  {pages} {084016} (\bibinfo {year} {2023})},\ \Eprint
  {http://arxiv.org/abs/2201.02220} {arXiv:2201.02220 [gr-qc]} \BibitemShut
  {NoStop}%
\bibitem [{\citenamefont {Detweiler}(1980)}]{Detweiler_1980}%
  \BibitemOpen
  \bibfield  {author} {\bibinfo {author} {\bibfnamefont {S.}~\bibnamefont
  {Detweiler}},\ }\href {\doibase 10.1103/PhysRevD.22.2323} {\bibfield
  {journal} {\bibinfo  {journal} {Phys. Rev. D}\ }\textbf {\bibinfo {volume}
  {22}},\ \bibinfo {pages} {2323} (\bibinfo {year} {1980})}\BibitemShut
  {NoStop}%
\bibitem [{\citenamefont {Dolan}(2007)}]{Dolan_2007}%
  \BibitemOpen
  \bibfield  {author} {\bibinfo {author} {\bibfnamefont {S.~R.}\ \bibnamefont
  {Dolan}},\ }\href {\doibase 10.1103/PhysRevD.76.084001} {\bibfield  {journal}
  {\bibinfo  {journal} {Phys. Rev. D}\ }\textbf {\bibinfo {volume} {76}},\
  \bibinfo {pages} {084001} (\bibinfo {year} {2007})}\BibitemShut {NoStop}%
\bibitem [{\citenamefont {Siemonsen}\ and\ \citenamefont
  {East}(2020)}]{Siemonsen_2020}%
  \BibitemOpen
  \bibfield  {author} {\bibinfo {author} {\bibfnamefont {N.}~\bibnamefont
  {Siemonsen}}\ and\ \bibinfo {author} {\bibfnamefont {W.~E.}\ \bibnamefont
  {East}},\ }\href {\doibase 10.1103/physrevd.101.024019} {\bibfield  {journal}
  {\bibinfo  {journal} {Physical Review D}\ }\textbf {\bibinfo {volume} {101}}
  (\bibinfo {year} {2020}),\ 10.1103/physrevd.101.024019}\BibitemShut {NoStop}%
\bibitem [{\citenamefont {Baryakhtar}\ \emph {et~al.}(2021)\citenamefont
  {Baryakhtar}, \citenamefont {Galanis}, \citenamefont {Lasenby},\ and\
  \citenamefont {Simon}}]{Baryakhtar_2021}%
  \BibitemOpen
  \bibfield  {author} {\bibinfo {author} {\bibfnamefont {M.}~\bibnamefont
  {Baryakhtar}}, \bibinfo {author} {\bibfnamefont {M.}~\bibnamefont {Galanis}},
  \bibinfo {author} {\bibfnamefont {R.}~\bibnamefont {Lasenby}}, \ and\
  \bibinfo {author} {\bibfnamefont {O.}~\bibnamefont {Simon}},\ }\href
  {\doibase 10.1103/PhysRevD.103.095019} {\bibfield  {journal} {\bibinfo
  {journal} {Phys. Rev. D}\ }\textbf {\bibinfo {volume} {103}},\ \bibinfo
  {pages} {095019} (\bibinfo {year} {2021})}\BibitemShut {NoStop}%
\bibitem [{\citenamefont {Ng}\ \emph {et~al.}(2021)\citenamefont {Ng},
  \citenamefont {Hannuksela}, \citenamefont {Vitale},\ and\ \citenamefont
  {Li}}]{Ng_2021}%
  \BibitemOpen
  \bibfield  {author} {\bibinfo {author} {\bibfnamefont {K.~K.~Y.}\
  \bibnamefont {Ng}}, \bibinfo {author} {\bibfnamefont {O.~A.}\ \bibnamefont
  {Hannuksela}}, \bibinfo {author} {\bibfnamefont {S.}~\bibnamefont {Vitale}},
  \ and\ \bibinfo {author} {\bibfnamefont {T.~G.~F.}\ \bibnamefont {Li}},\
  }\href {\doibase 10.1103/PhysRevD.103.063010} {\bibfield  {journal} {\bibinfo
   {journal} {Phys. Rev. D}\ }\textbf {\bibinfo {volume} {103}},\ \bibinfo
  {pages} {063010} (\bibinfo {year} {2021})}\BibitemShut {NoStop}%
\bibitem [{\citenamefont {Tong}\ \emph {et~al.}(2022)\citenamefont {Tong},
  \citenamefont {Wang},\ and\ \citenamefont {Zhu}}]{Tong_2022}%
  \BibitemOpen
  \bibfield  {author} {\bibinfo {author} {\bibfnamefont {X.}~\bibnamefont
  {Tong}}, \bibinfo {author} {\bibfnamefont {Y.}~\bibnamefont {Wang}}, \ and\
  \bibinfo {author} {\bibfnamefont {H.-Y.}\ \bibnamefont {Zhu}},\ }\href
  {\doibase 10.1103/physrevd.106.043002} {\bibfield  {journal} {\bibinfo
  {journal} {Physical Review D}\ }\textbf {\bibinfo {volume} {106}} (\bibinfo
  {year} {2022}),\ 10.1103/physrevd.106.043002}\BibitemShut {NoStop}%
\bibitem [{\citenamefont {Fan}\ \emph {et~al.}(2024)\citenamefont {Fan},
  \citenamefont {Tong}, \citenamefont {Wang},\ and\ \citenamefont
  {Zhu}}]{Fan_2024}%
  \BibitemOpen
  \bibfield  {author} {\bibinfo {author} {\bibfnamefont {K.}~\bibnamefont
  {Fan}}, \bibinfo {author} {\bibfnamefont {X.}~\bibnamefont {Tong}}, \bibinfo
  {author} {\bibfnamefont {Y.}~\bibnamefont {Wang}}, \ and\ \bibinfo {author}
  {\bibfnamefont {H.-Y.}\ \bibnamefont {Zhu}},\ }\href {\doibase
  10.1103/PhysRevD.109.024059} {\bibfield  {journal} {\bibinfo  {journal}
  {Phys. Rev. D}\ }\textbf {\bibinfo {volume} {109}},\ \bibinfo {pages}
  {024059} (\bibinfo {year} {2024})}\BibitemShut {NoStop}%
\bibitem [{\citenamefont {May}\ \emph {et~al.}(2025)\citenamefont {May},
  \citenamefont {East},\ and\ \citenamefont {Siemonsen}}]{May_2025}%
  \BibitemOpen
  \bibfield  {author} {\bibinfo {author} {\bibfnamefont {T.}~\bibnamefont
  {May}}, \bibinfo {author} {\bibfnamefont {W.~E.}\ \bibnamefont {East}}, \
  and\ \bibinfo {author} {\bibfnamefont {N.}~\bibnamefont {Siemonsen}},\ }\href
  {\doibase 10.1103/physrevd.111.044062} {\bibfield  {journal} {\bibinfo
  {journal} {Physical Review D}\ }\textbf {\bibinfo {volume} {111}} (\bibinfo
  {year} {2025}),\ 10.1103/physrevd.111.044062}\BibitemShut {NoStop}%
\bibitem [{\citenamefont {Arvanitaki}\ and\ \citenamefont
  {Dubovsky}(2011)}]{Arvanitaki_2011}%
  \BibitemOpen
  \bibfield  {author} {\bibinfo {author} {\bibfnamefont {A.}~\bibnamefont
  {Arvanitaki}}\ and\ \bibinfo {author} {\bibfnamefont {S.}~\bibnamefont
  {Dubovsky}},\ }\href {\doibase 10.1103/PhysRevD.83.044026} {\bibfield
  {journal} {\bibinfo  {journal} {Phys. Rev. D}\ }\textbf {\bibinfo {volume}
  {83}},\ \bibinfo {pages} {044026} (\bibinfo {year} {2011})}\BibitemShut
  {NoStop}%
\bibitem [{\citenamefont {Arvanitaki}\ \emph {et~al.}(2015)\citenamefont
  {Arvanitaki}, \citenamefont {Baryakhtar},\ and\ \citenamefont
  {Huang}}]{Arvanitaki_2015}%
  \BibitemOpen
  \bibfield  {author} {\bibinfo {author} {\bibfnamefont {A.}~\bibnamefont
  {Arvanitaki}}, \bibinfo {author} {\bibfnamefont {M.}~\bibnamefont
  {Baryakhtar}}, \ and\ \bibinfo {author} {\bibfnamefont {X.}~\bibnamefont
  {Huang}},\ }\href {\doibase 10.1103/PhysRevD.91.084011} {\bibfield  {journal}
  {\bibinfo  {journal} {Phys. Rev. D}\ }\textbf {\bibinfo {volume} {91}},\
  \bibinfo {pages} {084011} (\bibinfo {year} {2015})}\BibitemShut {NoStop}%
\bibitem [{\citenamefont {Berti}\ \emph {et~al.}(2019)\citenamefont {Berti},
  \citenamefont {Brito}, \citenamefont {Macedo}, \citenamefont {Raposo},\ and\
  \citenamefont {Rosa}}]{Berti:2019wnn}%
  \BibitemOpen
  \bibfield  {author} {\bibinfo {author} {\bibfnamefont {E.}~\bibnamefont
  {Berti}}, \bibinfo {author} {\bibfnamefont {R.}~\bibnamefont {Brito}},
  \bibinfo {author} {\bibfnamefont {C.~F.~B.}\ \bibnamefont {Macedo}}, \bibinfo
  {author} {\bibfnamefont {G.}~\bibnamefont {Raposo}}, \ and\ \bibinfo {author}
  {\bibfnamefont {J.~L.}\ \bibnamefont {Rosa}},\ }\href {\doibase
  10.1103/PhysRevD.99.104039} {\bibfield  {journal} {\bibinfo  {journal} {Phys.
  Rev. D}\ }\textbf {\bibinfo {volume} {99}},\ \bibinfo {pages} {104039}
  (\bibinfo {year} {2019})},\ \Eprint {http://arxiv.org/abs/1904.03131}
  {arXiv:1904.03131 [gr-qc]} \BibitemShut {NoStop}%
\bibitem [{\citenamefont {Baryakhtar}\ \emph {et~al.}(2017)\citenamefont
  {Baryakhtar}, \citenamefont {Lasenby},\ and\ \citenamefont
  {Teo}}]{Baryakhtar:2017ngi}%
  \BibitemOpen
  \bibfield  {author} {\bibinfo {author} {\bibfnamefont {M.}~\bibnamefont
  {Baryakhtar}}, \bibinfo {author} {\bibfnamefont {R.}~\bibnamefont {Lasenby}},
  \ and\ \bibinfo {author} {\bibfnamefont {M.}~\bibnamefont {Teo}},\ }\href
  {\doibase 10.1103/PhysRevD.96.035019} {\bibfield  {journal} {\bibinfo
  {journal} {Phys. Rev. D}\ }\textbf {\bibinfo {volume} {96}},\ \bibinfo
  {pages} {035019} (\bibinfo {year} {2017})},\ \Eprint
  {http://arxiv.org/abs/1704.05081} {arXiv:1704.05081 [hep-ph]} \BibitemShut
  {NoStop}%
\bibitem [{\citenamefont {Brito}\ \emph
  {et~al.}(2017{\natexlab{a}})\citenamefont {Brito}, \citenamefont {Ghosh},
  \citenamefont {Barausse}, \citenamefont {Berti}, \citenamefont {Cardoso},
  \citenamefont {Dvorkin}, \citenamefont {Klein},\ and\ \citenamefont
  {Pani}}]{Brito:2017wnc}%
  \BibitemOpen
  \bibfield  {author} {\bibinfo {author} {\bibfnamefont {R.}~\bibnamefont
  {Brito}}, \bibinfo {author} {\bibfnamefont {S.}~\bibnamefont {Ghosh}},
  \bibinfo {author} {\bibfnamefont {E.}~\bibnamefont {Barausse}}, \bibinfo
  {author} {\bibfnamefont {E.}~\bibnamefont {Berti}}, \bibinfo {author}
  {\bibfnamefont {V.}~\bibnamefont {Cardoso}}, \bibinfo {author} {\bibfnamefont
  {I.}~\bibnamefont {Dvorkin}}, \bibinfo {author} {\bibfnamefont
  {A.}~\bibnamefont {Klein}}, \ and\ \bibinfo {author} {\bibfnamefont
  {P.}~\bibnamefont {Pani}},\ }\href {\doibase 10.1103/PhysRevLett.119.131101}
  {\bibfield  {journal} {\bibinfo  {journal} {Phys. Rev. Lett.}\ }\textbf
  {\bibinfo {volume} {119}},\ \bibinfo {pages} {131101} (\bibinfo {year}
  {2017}{\natexlab{a}})},\ \Eprint {http://arxiv.org/abs/1706.05097}
  {arXiv:1706.05097 [gr-qc]} \BibitemShut {NoStop}%
\bibitem [{\citenamefont {Brito}\ \emph
  {et~al.}(2017{\natexlab{b}})\citenamefont {Brito}, \citenamefont {Ghosh},
  \citenamefont {Barausse}, \citenamefont {Berti}, \citenamefont {Cardoso},
  \citenamefont {Dvorkin}, \citenamefont {Klein},\ and\ \citenamefont
  {Pani}}]{Brito:2017zvb}%
  \BibitemOpen
  \bibfield  {author} {\bibinfo {author} {\bibfnamefont {R.}~\bibnamefont
  {Brito}}, \bibinfo {author} {\bibfnamefont {S.}~\bibnamefont {Ghosh}},
  \bibinfo {author} {\bibfnamefont {E.}~\bibnamefont {Barausse}}, \bibinfo
  {author} {\bibfnamefont {E.}~\bibnamefont {Berti}}, \bibinfo {author}
  {\bibfnamefont {V.}~\bibnamefont {Cardoso}}, \bibinfo {author} {\bibfnamefont
  {I.}~\bibnamefont {Dvorkin}}, \bibinfo {author} {\bibfnamefont
  {A.}~\bibnamefont {Klein}}, \ and\ \bibinfo {author} {\bibfnamefont
  {P.}~\bibnamefont {Pani}},\ }\href {\doibase 10.1103/PhysRevD.96.064050}
  {\bibfield  {journal} {\bibinfo  {journal} {Phys. Rev. D}\ }\textbf {\bibinfo
  {volume} {96}},\ \bibinfo {pages} {064050} (\bibinfo {year}
  {2017}{\natexlab{b}})},\ \Eprint {http://arxiv.org/abs/1706.06311}
  {arXiv:1706.06311 [gr-qc]} \BibitemShut {NoStop}%
\bibitem [{\citenamefont {D'Antonio}\ \emph {et~al.}(2018)\citenamefont
  {D'Antonio} \emph {et~al.}}]{DAntonio:2018sff}%
  \BibitemOpen
  \bibfield  {author} {\bibinfo {author} {\bibfnamefont {S.}~\bibnamefont
  {D'Antonio}} \emph {et~al.},\ }\href {\doibase 10.1103/PhysRevD.98.103017}
  {\bibfield  {journal} {\bibinfo  {journal} {Phys. Rev. D}\ }\textbf {\bibinfo
  {volume} {98}},\ \bibinfo {pages} {103017} (\bibinfo {year} {2018})},\
  \Eprint {http://arxiv.org/abs/1809.07202} {arXiv:1809.07202 [gr-qc]}
  \BibitemShut {NoStop}%
\bibitem [{\citenamefont {Ghosh}\ \emph {et~al.}(2019)\citenamefont {Ghosh},
  \citenamefont {Berti}, \citenamefont {Brito},\ and\ \citenamefont
  {Richartz}}]{Ghosh:2018gaw}%
  \BibitemOpen
  \bibfield  {author} {\bibinfo {author} {\bibfnamefont {S.}~\bibnamefont
  {Ghosh}}, \bibinfo {author} {\bibfnamefont {E.}~\bibnamefont {Berti}},
  \bibinfo {author} {\bibfnamefont {R.}~\bibnamefont {Brito}}, \ and\ \bibinfo
  {author} {\bibfnamefont {M.}~\bibnamefont {Richartz}},\ }\href {\doibase
  10.1103/PhysRevD.99.104030} {\bibfield  {journal} {\bibinfo  {journal} {Phys.
  Rev. D}\ }\textbf {\bibinfo {volume} {99}},\ \bibinfo {pages} {104030}
  (\bibinfo {year} {2019})},\ \Eprint {http://arxiv.org/abs/1812.01620}
  {arXiv:1812.01620 [gr-qc]} \BibitemShut {NoStop}%
\bibitem [{\citenamefont {Xie}\ \emph {et~al.}(2025)\citenamefont {Xie},
  \citenamefont {Chung}, \citenamefont {Sotiriou},\ and\ \citenamefont
  {Yunes}}]{Xie_2025}%
  \BibitemOpen
  \bibfield  {author} {\bibinfo {author} {\bibfnamefont {Y.}~\bibnamefont
  {Xie}}, \bibinfo {author} {\bibfnamefont {A.~K.-W.}\ \bibnamefont {Chung}},
  \bibinfo {author} {\bibfnamefont {T.~P.}\ \bibnamefont {Sotiriou}}, \ and\
  \bibinfo {author} {\bibfnamefont {N.}~\bibnamefont {Yunes}},\ }\href
  {\doibase 10.1103/PhysRevLett.134.191402} {\bibfield  {journal} {\bibinfo
  {journal} {Phys. Rev. Lett.}\ }\textbf {\bibinfo {volume} {134}},\ \bibinfo
  {pages} {191402} (\bibinfo {year} {2025})}\BibitemShut {NoStop}%
\bibitem [{\citenamefont {Maselli}\ \emph {et~al.}(2022)\citenamefont
  {Maselli}, \citenamefont {Franchini}, \citenamefont {Gualtieri},
  \citenamefont {Sotiriou}, \citenamefont {Barsanti},\ and\ \citenamefont
  {Pani}}]{Maselli_2021}%
  \BibitemOpen
  \bibfield  {author} {\bibinfo {author} {\bibfnamefont {A.}~\bibnamefont
  {Maselli}}, \bibinfo {author} {\bibfnamefont {N.}~\bibnamefont {Franchini}},
  \bibinfo {author} {\bibfnamefont {L.}~\bibnamefont {Gualtieri}}, \bibinfo
  {author} {\bibfnamefont {T.~P.}\ \bibnamefont {Sotiriou}}, \bibinfo {author}
  {\bibfnamefont {S.}~\bibnamefont {Barsanti}}, \ and\ \bibinfo {author}
  {\bibfnamefont {P.}~\bibnamefont {Pani}},\ }\href {\doibase
  10.1038/s41550-021-01589-5} {\bibfield  {journal} {\bibinfo  {journal}
  {Nature Astron.}\ }\textbf {\bibinfo {volume} {6}},\ \bibinfo {pages} {464}
  (\bibinfo {year} {2022})},\ \Eprint {http://arxiv.org/abs/2106.11325}
  {arXiv:2106.11325 [gr-qc]} \BibitemShut {NoStop}%
\bibitem [{\citenamefont {Duque}\ \emph {et~al.}(2024)\citenamefont {Duque},
  \citenamefont {Macedo}, \citenamefont {Vicente},\ and\ \citenamefont
  {Cardoso}}]{Duque_2024}%
  \BibitemOpen
  \bibfield  {author} {\bibinfo {author} {\bibfnamefont {F.}~\bibnamefont
  {Duque}}, \bibinfo {author} {\bibfnamefont {C.~F.}\ \bibnamefont {Macedo}},
  \bibinfo {author} {\bibfnamefont {R.}~\bibnamefont {Vicente}}, \ and\
  \bibinfo {author} {\bibfnamefont {V.}~\bibnamefont {Cardoso}},\ }\href
  {\doibase 10.1103/physrevlett.133.121404} {\bibfield  {journal} {\bibinfo
  {journal} {Physical Review Letters}\ }\textbf {\bibinfo {volume} {133}}
  (\bibinfo {year} {2024}),\ 10.1103/physrevlett.133.121404}\BibitemShut
  {NoStop}%
\bibitem [{\citenamefont {Wade}\ and\ \citenamefont
  {Nichols}(2026)}]{Wade_2026}%
  \BibitemOpen
  \bibfield  {author} {\bibinfo {author} {\bibfnamefont {B.~A.}\ \bibnamefont
  {Wade}}\ and\ \bibinfo {author} {\bibfnamefont {D.~A.}\ \bibnamefont
  {Nichols}},\ }\href {\doibase 10.1103/5s63-kgx2} {\bibfield  {journal}
  {\bibinfo  {journal} {Phys. Rev. D}\ }\textbf {\bibinfo {volume} {113}},\
  \bibinfo {pages} {044014} (\bibinfo {year} {2026})}\BibitemShut {NoStop}%
\bibitem [{\citenamefont {Rahman}\ and\ \citenamefont
  {Takahashi}(2026)}]{Rahman_2026}%
  \BibitemOpen
  \bibfield  {author} {\bibinfo {author} {\bibfnamefont {M.}~\bibnamefont
  {Rahman}}\ and\ \bibinfo {author} {\bibfnamefont {T.}~\bibnamefont
  {Takahashi}},\ }\href {\doibase 10.1103/76m9-rpzj} {\bibfield  {journal}
  {\bibinfo  {journal} {Phys. Rev. D}\ }\textbf {\bibinfo {volume} {113}},\
  \bibinfo {pages} {044033} (\bibinfo {year} {2026})}\BibitemShut {NoStop}%
\bibitem [{\citenamefont {Lui}\ \emph {et~al.}(2026)\citenamefont {Lui},
  \citenamefont {Chung},\ and\ \citenamefont {Torres-Orjuela}}]{Lui_2026c}%
  \BibitemOpen
  \bibfield  {author} {\bibinfo {author} {\bibfnamefont {L.}~\bibnamefont
  {Lui}}, \bibinfo {author} {\bibfnamefont {A.~K.-W.}\ \bibnamefont {Chung}}, \
  and\ \bibinfo {author} {\bibfnamefont {A.}~\bibnamefont {Torres-Orjuela}},\
  }\href {https://arxiv.org/abs/2608.28037} {\enquote {\bibinfo {title}
  {Devoured by a hairy gargantua: Probing massive scalar charges with
  non-minimal curvature coupling with extreme-mass-ratio inspirals},}\ }
  (\bibinfo {year} {2026}),\ \Eprint {http://arxiv.org/abs/2608.28037}
  {arXiv:2608.28037 [gr-qc]} \BibitemShut {NoStop}%
\bibitem [{\citenamefont {Li}\ \emph {et~al.}(2026{\natexlab{a}})\citenamefont
  {Li}, \citenamefont {Kuan}, \citenamefont {Weller}, \citenamefont {Yang},
  \citenamefont {Lam}, \citenamefont {Bourg}, \citenamefont {Chung},
  \citenamefont {Chen}, \citenamefont {Yunes},\ and\ \citenamefont
  {Yang}}]{Li:2026foy}%
  \BibitemOpen
  \bibfield  {author} {\bibinfo {author} {\bibfnamefont {D.}~\bibnamefont
  {Li}}, \bibinfo {author} {\bibfnamefont {H.-J.}\ \bibnamefont {Kuan}},
  \bibinfo {author} {\bibfnamefont {C.}~\bibnamefont {Weller}}, \bibinfo
  {author} {\bibfnamefont {C.}~\bibnamefont {Yang}}, \bibinfo {author}
  {\bibfnamefont {K.~K.-H.}\ \bibnamefont {Lam}}, \bibinfo {author}
  {\bibfnamefont {P.}~\bibnamefont {Bourg}}, \bibinfo {author} {\bibfnamefont
  {A.~K.-W.}\ \bibnamefont {Chung}}, \bibinfo {author} {\bibfnamefont
  {Y.}~\bibnamefont {Chen}}, \bibinfo {author} {\bibfnamefont {N.}~\bibnamefont
  {Yunes}}, \ and\ \bibinfo {author} {\bibfnamefont {H.}~\bibnamefont {Yang}},\
  }\href@noop {} {\  (\bibinfo {year} {2026}{\natexlab{a}})},\ \Eprint
  {http://arxiv.org/abs/2610.22480} {arXiv:2610.22480 [gr-qc]} \BibitemShut
  {NoStop}%
\bibitem [{\citenamefont {Tsukada}\ \emph {et~al.}(2019)\citenamefont
  {Tsukada}, \citenamefont {Callister}, \citenamefont {Matas},\ and\
  \citenamefont {Meyers}}]{Tsukada:2018mbp}%
  \BibitemOpen
  \bibfield  {author} {\bibinfo {author} {\bibfnamefont {L.}~\bibnamefont
  {Tsukada}}, \bibinfo {author} {\bibfnamefont {T.}~\bibnamefont {Callister}},
  \bibinfo {author} {\bibfnamefont {A.}~\bibnamefont {Matas}}, \ and\ \bibinfo
  {author} {\bibfnamefont {P.}~\bibnamefont {Meyers}},\ }\href {\doibase
  10.1103/PhysRevD.99.103015} {\bibfield  {journal} {\bibinfo  {journal} {Phys.
  Rev. D}\ }\textbf {\bibinfo {volume} {99}},\ \bibinfo {pages} {103015}
  (\bibinfo {year} {2019})},\ \Eprint {http://arxiv.org/abs/1812.09622}
  {arXiv:1812.09622 [astro-ph.HE]} \BibitemShut {NoStop}%
\bibitem [{\citenamefont {Tsukada}\ \emph {et~al.}(2021)\citenamefont
  {Tsukada}, \citenamefont {Brito}, \citenamefont {East},\ and\ \citenamefont
  {Siemonsen}}]{Tsukada:2020lgt}%
  \BibitemOpen
  \bibfield  {author} {\bibinfo {author} {\bibfnamefont {L.}~\bibnamefont
  {Tsukada}}, \bibinfo {author} {\bibfnamefont {R.}~\bibnamefont {Brito}},
  \bibinfo {author} {\bibfnamefont {W.~E.}\ \bibnamefont {East}}, \ and\
  \bibinfo {author} {\bibfnamefont {N.}~\bibnamefont {Siemonsen}},\ }\href
  {\doibase 10.1103/PhysRevD.103.083005} {\bibfield  {journal} {\bibinfo
  {journal} {Phys. Rev. D}\ }\textbf {\bibinfo {volume} {103}},\ \bibinfo
  {pages} {083005} (\bibinfo {year} {2021})},\ \Eprint
  {http://arxiv.org/abs/2011.06995} {arXiv:2011.06995 [astro-ph.HE]}
  \BibitemShut {NoStop}%
\bibitem [{\citenamefont {Cano}\ and\ \citenamefont
  {Ruip{\'e}rez}(2019)}]{Cano:2019ore}%
  \BibitemOpen
  \bibfield  {author} {\bibinfo {author} {\bibfnamefont {P.~A.}\ \bibnamefont
  {Cano}}\ and\ \bibinfo {author} {\bibfnamefont {A.}~\bibnamefont
  {Ruip{\'e}rez}},\ }\href {\doibase 10.1007/JHEP05(2019)189} {\bibfield
  {journal} {\bibinfo  {journal} {JHEP}\ }\textbf {\bibinfo {volume} {05}},\
  \bibinfo {pages} {189} (\bibinfo {year} {2019})},\ \bibinfo {note} {[Erratum:
  JHEP 03, 187 (2020)]},\ \Eprint {http://arxiv.org/abs/1901.01315}
  {arXiv:1901.01315 [gr-qc]} \BibitemShut {NoStop}%
\bibitem [{\citenamefont {Cano}\ and\ \citenamefont
  {Ruip{\'e}rez}(2022)}]{Cano:2021rey}%
  \BibitemOpen
  \bibfield  {author} {\bibinfo {author} {\bibfnamefont {P.~A.}\ \bibnamefont
  {Cano}}\ and\ \bibinfo {author} {\bibfnamefont {A.}~\bibnamefont
  {Ruip{\'e}rez}},\ }\href {\doibase 10.1103/PhysRevD.105.044022} {\bibfield
  {journal} {\bibinfo  {journal} {Phys. Rev. D}\ }\textbf {\bibinfo {volume}
  {105}},\ \bibinfo {pages} {044022} (\bibinfo {year} {2022})},\ \Eprint
  {http://arxiv.org/abs/2111.04750} {arXiv:2111.04750 [hep-th]} \BibitemShut
  {NoStop}%
\bibitem [{\citenamefont {Alexander}\ and\ \citenamefont
  {Yunes}(2009)}]{Alexander_2009}%
  \BibitemOpen
  \bibfield  {author} {\bibinfo {author} {\bibfnamefont {S.}~\bibnamefont
  {Alexander}}\ and\ \bibinfo {author} {\bibfnamefont {N.}~\bibnamefont
  {Yunes}},\ }\href {\doibase 10.1016/j.physrep.2009.07.002} {\bibfield
  {journal} {\bibinfo  {journal} {Phys. Rept.}\ }\textbf {\bibinfo {volume}
  {480}},\ \bibinfo {pages} {1} (\bibinfo {year} {2009})},\ \Eprint
  {http://arxiv.org/abs/0907.2562} {arXiv:0907.2562 [hep-th]} \BibitemShut
  {NoStop}%
\bibitem [{\citenamefont {Jackiw}\ and\ \citenamefont
  {Pi}(2003)}]{Jackiw:2003pm}%
  \BibitemOpen
  \bibfield  {author} {\bibinfo {author} {\bibfnamefont {R.}~\bibnamefont
  {Jackiw}}\ and\ \bibinfo {author} {\bibfnamefont {S.~Y.}\ \bibnamefont
  {Pi}},\ }\href {\doibase 10.1103/PhysRevD.68.104012} {\bibfield  {journal}
  {\bibinfo  {journal} {Phys. Rev. D}\ }\textbf {\bibinfo {volume} {68}},\
  \bibinfo {pages} {104012} (\bibinfo {year} {2003})},\ \Eprint
  {http://arxiv.org/abs/gr-qc/0308071} {arXiv:gr-qc/0308071} \BibitemShut
  {NoStop}%
\bibitem [{\citenamefont {Lam}\ \emph {et~al.}(2026{\natexlab{a}})\citenamefont
  {Lam}, \citenamefont {Chung},\ and\ \citenamefont {Yunes}}]{Lam_2026}%
  \BibitemOpen
  \bibfield  {author} {\bibinfo {author} {\bibfnamefont {K.~K.-H.}\
  \bibnamefont {Lam}}, \bibinfo {author} {\bibfnamefont {A.~K.-W.}\
  \bibnamefont {Chung}}, \ and\ \bibinfo {author} {\bibfnamefont
  {N.}~\bibnamefont {Yunes}},\ }\href {\doibase 10.1103/6z43-xdv7} {\bibfield
  {journal} {\bibinfo  {journal} {Phys. Rev. Lett.}\ }\textbf {\bibinfo
  {volume} {136}},\ \bibinfo {pages} {021401} (\bibinfo {year}
  {2026}{\natexlab{a}})}\BibitemShut {NoStop}%
\bibitem [{\citenamefont {Lam}\ \emph {et~al.}(2026{\natexlab{b}})\citenamefont
  {Lam}, \citenamefont {Chung},\ and\ \citenamefont {Yunes}}]{Lam:2025fzi}%
  \BibitemOpen
  \bibfield  {author} {\bibinfo {author} {\bibfnamefont {K.~K.-H.}\
  \bibnamefont {Lam}}, \bibinfo {author} {\bibfnamefont {A.~K.-W.}\
  \bibnamefont {Chung}}, \ and\ \bibinfo {author} {\bibfnamefont
  {N.}~\bibnamefont {Yunes}},\ }\href {\doibase 10.1103/txxb-z8bx} {\bibfield
  {journal} {\bibinfo  {journal} {Phys. Rev. D}\ }\textbf {\bibinfo {volume}
  {113}},\ \bibinfo {pages} {024030} (\bibinfo {year} {2026}{\natexlab{b}})},\
  \Eprint {http://arxiv.org/abs/2510.05208} {arXiv:2510.05208 [gr-qc]}
  \BibitemShut {NoStop}%
\bibitem [{\citenamefont {Chung}(2026)}]{Chung_2026b}%
  \BibitemOpen
  \bibfield  {author} {\bibinfo {author} {\bibfnamefont {A.~K.-W.}\
  \bibnamefont {Chung}},\ }\href {\doibase 10.1103/4qbd-f98r} {\bibfield
  {journal} {\bibinfo  {journal} {Phys. Rev. D}\ }\textbf {\bibinfo {volume}
  {113}},\ \bibinfo {pages} {095032} (\bibinfo {year} {2026})}\BibitemShut
  {NoStop}%
\bibitem [{\citenamefont {Li}\ \emph {et~al.}(2026{\natexlab{b}})\citenamefont
  {Li}, \citenamefont {Hu},\ and\ \citenamefont {Yunes}}]{Li:2026hpw}%
  \BibitemOpen
  \bibfield  {author} {\bibinfo {author} {\bibfnamefont {D.}~\bibnamefont
  {Li}}, \bibinfo {author} {\bibfnamefont {J.}~\bibnamefont {Hu}}, \ and\
  \bibinfo {author} {\bibfnamefont {N.}~\bibnamefont {Yunes}},\ }\href@noop {}
  {\  (\bibinfo {year} {2026}{\natexlab{b}})},\ \Eprint
  {http://arxiv.org/abs/2610.37515} {arXiv:2610.37515 [gr-qc]} \BibitemShut
  {NoStop}%
\bibitem [{\citenamefont {Chung}\ \emph {et~al.}(2023)\citenamefont {Chung},
  \citenamefont {Wagle},\ and\ \citenamefont {Yunes}}]{Chung_2023}%
  \BibitemOpen
  \bibfield  {author} {\bibinfo {author} {\bibfnamefont {A.~K.-W.}\
  \bibnamefont {Chung}}, \bibinfo {author} {\bibfnamefont {P.}~\bibnamefont
  {Wagle}}, \ and\ \bibinfo {author} {\bibfnamefont {N.}~\bibnamefont
  {Yunes}},\ }\href {\doibase 10.1103/PhysRevD.107.124032} {\bibfield
  {journal} {\bibinfo  {journal} {Phys. Rev. D}\ }\textbf {\bibinfo {volume}
  {107}},\ \bibinfo {pages} {124032} (\bibinfo {year} {2023})}\BibitemShut
  {NoStop}%
\bibitem [{\citenamefont {Chung}\ and\ \citenamefont
  {Yunes}(2024{\natexlab{a}})}]{Chung_2024b}%
  \BibitemOpen
  \bibfield  {author} {\bibinfo {author} {\bibfnamefont {A.~K.-W.}\
  \bibnamefont {Chung}}\ and\ \bibinfo {author} {\bibfnamefont
  {N.}~\bibnamefont {Yunes}},\ }\href {\doibase 10.1103/PhysRevLett.133.181401}
  {\bibfield  {journal} {\bibinfo  {journal} {Phys. Rev. Lett.}\ }\textbf
  {\bibinfo {volume} {133}},\ \bibinfo {pages} {181401} (\bibinfo {year}
  {2024}{\natexlab{a}})}\BibitemShut {NoStop}%
\bibitem [{\citenamefont {Chung}\ and\ \citenamefont
  {Yunes}(2024{\natexlab{b}})}]{Chung:2024vaf}%
  \BibitemOpen
  \bibfield  {author} {\bibinfo {author} {\bibfnamefont {A.~K.-W.}\
  \bibnamefont {Chung}}\ and\ \bibinfo {author} {\bibfnamefont
  {N.}~\bibnamefont {Yunes}},\ }\href {\doibase 10.1103/PhysRevD.110.064019}
  {\bibfield  {journal} {\bibinfo  {journal} {Phys. Rev. D}\ }\textbf {\bibinfo
  {volume} {110}},\ \bibinfo {pages} {064019} (\bibinfo {year}
  {2024}{\natexlab{b}})},\ \Eprint {http://arxiv.org/abs/2406.11986}
  {arXiv:2406.11986 [gr-qc]} \BibitemShut {NoStop}%
\bibitem [{\citenamefont {Chung}\ \emph {et~al.}(2025)\citenamefont {Chung},
  \citenamefont {Lam},\ and\ \citenamefont {Yunes}}]{Chung:2025gyg}%
  \BibitemOpen
  \bibfield  {author} {\bibinfo {author} {\bibfnamefont {A.~K.-W.}\
  \bibnamefont {Chung}}, \bibinfo {author} {\bibfnamefont {K.~K.-H.}\
  \bibnamefont {Lam}}, \ and\ \bibinfo {author} {\bibfnamefont
  {N.}~\bibnamefont {Yunes}},\ }\href {\doibase 10.1103/g83n-rrlj} {\bibfield
  {journal} {\bibinfo  {journal} {Phys. Rev. D}\ }\textbf {\bibinfo {volume}
  {111}},\ \bibinfo {pages} {124052} (\bibinfo {year} {2025})},\ \Eprint
  {http://arxiv.org/abs/2503.11759} {arXiv:2503.11759 [gr-qc]} \BibitemShut
  {NoStop}%
\bibitem [{\citenamefont {Li}\ \emph {et~al.}(2023)\citenamefont {Li},
  \citenamefont {Wagle}, \citenamefont {Chen},\ and\ \citenamefont
  {Yunes}}]{Li:2022pcy}%
  \BibitemOpen
  \bibfield  {author} {\bibinfo {author} {\bibfnamefont {D.}~\bibnamefont
  {Li}}, \bibinfo {author} {\bibfnamefont {P.}~\bibnamefont {Wagle}}, \bibinfo
  {author} {\bibfnamefont {Y.}~\bibnamefont {Chen}}, \ and\ \bibinfo {author}
  {\bibfnamefont {N.}~\bibnamefont {Yunes}},\ }\href {\doibase
  10.1103/PhysRevX.13.021029} {\bibfield  {journal} {\bibinfo  {journal} {Phys.
  Rev. X}\ }\textbf {\bibinfo {volume} {13}},\ \bibinfo {pages} {021029}
  (\bibinfo {year} {2023})},\ \Eprint {http://arxiv.org/abs/2206.10652}
  {arXiv:2206.10652 [gr-qc]} \BibitemShut {NoStop}%
\bibitem [{\citenamefont {Li}\ \emph {et~al.}(2024)\citenamefont {Li},
  \citenamefont {Hussain}, \citenamefont {Wagle}, \citenamefont {Chen},
  \citenamefont {Yunes},\ and\ \citenamefont {Zimmerman}}]{Li:2023ulk}%
  \BibitemOpen
  \bibfield  {author} {\bibinfo {author} {\bibfnamefont {D.}~\bibnamefont
  {Li}}, \bibinfo {author} {\bibfnamefont {A.}~\bibnamefont {Hussain}},
  \bibinfo {author} {\bibfnamefont {P.}~\bibnamefont {Wagle}}, \bibinfo
  {author} {\bibfnamefont {Y.}~\bibnamefont {Chen}}, \bibinfo {author}
  {\bibfnamefont {N.}~\bibnamefont {Yunes}}, \ and\ \bibinfo {author}
  {\bibfnamefont {A.}~\bibnamefont {Zimmerman}},\ }\href {\doibase
  10.1103/PhysRevD.109.104026} {\bibfield  {journal} {\bibinfo  {journal}
  {Phys. Rev. D}\ }\textbf {\bibinfo {volume} {109}},\ \bibinfo {pages}
  {104026} (\bibinfo {year} {2024})},\ \Eprint
  {http://arxiv.org/abs/2310.06033} {arXiv:2310.06033 [gr-qc]} \BibitemShut
  {NoStop}%
\bibitem [{\citenamefont {Li}\ \emph {et~al.}(2025)\citenamefont {Li},
  \citenamefont {Wagle}, \citenamefont {Chen},\ and\ \citenamefont
  {Yunes}}]{Li:2025fci}%
  \BibitemOpen
  \bibfield  {author} {\bibinfo {author} {\bibfnamefont {D.}~\bibnamefont
  {Li}}, \bibinfo {author} {\bibfnamefont {P.}~\bibnamefont {Wagle}}, \bibinfo
  {author} {\bibfnamefont {Y.}~\bibnamefont {Chen}}, \ and\ \bibinfo {author}
  {\bibfnamefont {N.}~\bibnamefont {Yunes}},\ }\href {\doibase
  10.1103/yfw6-32yl} {\bibfield  {journal} {\bibinfo  {journal} {Phys. Rev. D}\
  }\textbf {\bibinfo {volume} {112}},\ \bibinfo {pages} {044005} (\bibinfo
  {year} {2025})},\ \Eprint {http://arxiv.org/abs/2503.15606} {arXiv:2503.15606
  [gr-qc]} \BibitemShut {NoStop}%
\bibitem [{\citenamefont {Weller}\ \emph {et~al.}(2026)\citenamefont {Weller},
  \citenamefont {Li}, \citenamefont {Wagle}, \citenamefont {Laeuger},
  \citenamefont {Chen},\ and\ \citenamefont {Yunes}}]{Weller:2026jqu}%
  \BibitemOpen
  \bibfield  {author} {\bibinfo {author} {\bibfnamefont {C.}~\bibnamefont
  {Weller}}, \bibinfo {author} {\bibfnamefont {D.}~\bibnamefont {Li}}, \bibinfo
  {author} {\bibfnamefont {P.}~\bibnamefont {Wagle}}, \bibinfo {author}
  {\bibfnamefont {A.}~\bibnamefont {Laeuger}}, \bibinfo {author} {\bibfnamefont
  {Y.}~\bibnamefont {Chen}}, \ and\ \bibinfo {author} {\bibfnamefont
  {N.}~\bibnamefont {Yunes}},\ }\href@noop {} {\  (\bibinfo {year} {2026})},\
  \Eprint {http://arxiv.org/abs/2608.21532} {arXiv:2608.21532 [gr-qc]}
  \BibitemShut {NoStop}%
\bibitem [{\citenamefont {Bl{\'a}zquez-Salcedo}\ \emph
  {et~al.}(2024)\citenamefont {Bl{\'a}zquez-Salcedo}, \citenamefont {Khoo},
  \citenamefont {Kunz},\ and\ \citenamefont
  {Gonz{\'a}lez-Romero}}]{Blazquez-Salcedo:2023hwg}%
  \BibitemOpen
  \bibfield  {author} {\bibinfo {author} {\bibfnamefont {J.~L.}\ \bibnamefont
  {Bl{\'a}zquez-Salcedo}}, \bibinfo {author} {\bibfnamefont {F.~S.}\
  \bibnamefont {Khoo}}, \bibinfo {author} {\bibfnamefont {J.}~\bibnamefont
  {Kunz}}, \ and\ \bibinfo {author} {\bibfnamefont {L.~M.}\ \bibnamefont
  {Gonz{\'a}lez-Romero}},\ }\href {\doibase 10.1103/PhysRevD.109.064028}
  {\bibfield  {journal} {\bibinfo  {journal} {Phys. Rev. D}\ }\textbf {\bibinfo
  {volume} {109}},\ \bibinfo {pages} {064028} (\bibinfo {year} {2024})},\
  \Eprint {http://arxiv.org/abs/2312.10754} {arXiv:2312.10754 [gr-qc]}
  \BibitemShut {NoStop}%
\bibitem [{\citenamefont {Bl{\'a}zquez-Salcedo}\ \emph
  {et~al.}(2025)\citenamefont {Bl{\'a}zquez-Salcedo}, \citenamefont {Khoo},
  \citenamefont {Kleihaus},\ and\ \citenamefont
  {Kunz}}]{Blazquez-Salcedo:2024oek}%
  \BibitemOpen
  \bibfield  {author} {\bibinfo {author} {\bibfnamefont {J.~L.}\ \bibnamefont
  {Bl{\'a}zquez-Salcedo}}, \bibinfo {author} {\bibfnamefont {F.~S.}\
  \bibnamefont {Khoo}}, \bibinfo {author} {\bibfnamefont {B.}~\bibnamefont
  {Kleihaus}}, \ and\ \bibinfo {author} {\bibfnamefont {J.}~\bibnamefont
  {Kunz}},\ }\href {\doibase 10.1103/PhysRevD.111.L021505} {\bibfield
  {journal} {\bibinfo  {journal} {Phys. Rev. D}\ }\textbf {\bibinfo {volume}
  {111}},\ \bibinfo {pages} {L021505} (\bibinfo {year} {2025})},\ \Eprint
  {http://arxiv.org/abs/2407.20760} {arXiv:2407.20760 [gr-qc]} \BibitemShut
  {NoStop}%
\bibitem [{\citenamefont {Khoo}\ \emph {et~al.}(2025)\citenamefont {Khoo},
  \citenamefont {Bl{\'a}zquez-Salcedo}, \citenamefont {Kleihaus},\ and\
  \citenamefont {Kunz}}]{Khoo:2024agm}%
  \BibitemOpen
  \bibfield  {author} {\bibinfo {author} {\bibfnamefont {F.~S.}\ \bibnamefont
  {Khoo}}, \bibinfo {author} {\bibfnamefont {J.~L.}\ \bibnamefont
  {Bl{\'a}zquez-Salcedo}}, \bibinfo {author} {\bibfnamefont {B.}~\bibnamefont
  {Kleihaus}}, \ and\ \bibinfo {author} {\bibfnamefont {J.}~\bibnamefont
  {Kunz}},\ }\href {\doibase 10.1140/epjc/s10052-025-15106-9} {\bibfield
  {journal} {\bibinfo  {journal} {Eur. Phys. J. C}\ }\textbf {\bibinfo {volume}
  {85}},\ \bibinfo {pages} {1366} (\bibinfo {year} {2025})},\ \Eprint
  {http://arxiv.org/abs/2412.09377} {arXiv:2412.09377 [gr-qc]} \BibitemShut
  {NoStop}%
\bibitem [{\citenamefont {Khoo}\ \emph {et~al.}(2024)\citenamefont {Khoo},
  \citenamefont {Azad}, \citenamefont {Bl{\'a}zquez-Salcedo}, \citenamefont
  {Gonz{\'a}lez-Romero}, \citenamefont {Kleihaus}, \citenamefont {Kunz},\ and\
  \citenamefont {Navarro-L{\'e}rida}}]{Khoo:2024yeh}%
  \BibitemOpen
  \bibfield  {author} {\bibinfo {author} {\bibfnamefont {F.~S.}\ \bibnamefont
  {Khoo}}, \bibinfo {author} {\bibfnamefont {B.}~\bibnamefont {Azad}}, \bibinfo
  {author} {\bibfnamefont {J.~L.}\ \bibnamefont {Bl{\'a}zquez-Salcedo}},
  \bibinfo {author} {\bibfnamefont {L.~M.}\ \bibnamefont
  {Gonz{\'a}lez-Romero}}, \bibinfo {author} {\bibfnamefont {B.}~\bibnamefont
  {Kleihaus}}, \bibinfo {author} {\bibfnamefont {J.}~\bibnamefont {Kunz}}, \
  and\ \bibinfo {author} {\bibfnamefont {F.}~\bibnamefont
  {Navarro-L{\'e}rida}},\ }\href {\doibase 10.1103/PhysRevD.109.084013}
  {\bibfield  {journal} {\bibinfo  {journal} {Phys. Rev. D}\ }\textbf {\bibinfo
  {volume} {109}},\ \bibinfo {pages} {084013} (\bibinfo {year} {2024})},\
  \Eprint {http://arxiv.org/abs/2401.02898} {arXiv:2401.02898 [gr-qc]}
  \BibitemShut {NoStop}%
\bibitem [{\citenamefont {Cano}\ \emph {et~al.}(2023)\citenamefont {Cano},
  \citenamefont {Fransen}, \citenamefont {Hertog},\ and\ \citenamefont
  {Maenaut}}]{Cano:2023jbk}%
  \BibitemOpen
  \bibfield  {author} {\bibinfo {author} {\bibfnamefont {P.~A.}\ \bibnamefont
  {Cano}}, \bibinfo {author} {\bibfnamefont {K.}~\bibnamefont {Fransen}},
  \bibinfo {author} {\bibfnamefont {T.}~\bibnamefont {Hertog}}, \ and\ \bibinfo
  {author} {\bibfnamefont {S.}~\bibnamefont {Maenaut}},\ }\href {\doibase
  10.1103/PhysRevD.108.124032} {\bibfield  {journal} {\bibinfo  {journal}
  {Phys. Rev. D}\ }\textbf {\bibinfo {volume} {108}},\ \bibinfo {pages}
  {124032} (\bibinfo {year} {2023})},\ \Eprint
  {http://arxiv.org/abs/2307.07431} {arXiv:2307.07431 [gr-qc]} \BibitemShut
  {NoStop}%
\bibitem [{\citenamefont {Lui}\ and\ \citenamefont
  {Chung}(2026)}]{Lui:2026osx}%
  \BibitemOpen
  \bibfield  {author} {\bibinfo {author} {\bibfnamefont {L.}~\bibnamefont
  {Lui}}\ and\ \bibinfo {author} {\bibfnamefont {A.~K.-W.}\ \bibnamefont
  {Chung}},\ }\href@noop {} {\  (\bibinfo {year} {2026})},\ \Eprint
  {http://arxiv.org/abs/2610.16054} {arXiv:2610.16054 [gr-qc]} \BibitemShut
  {NoStop}%
\bibitem [{\citenamefont {Chung}\ and\ \citenamefont
  {Yunes}(2026)}]{Chung:2025ucb}%
  \BibitemOpen
  \bibfield  {author} {\bibinfo {author} {\bibfnamefont {A.~K.-W.}\
  \bibnamefont {Chung}}\ and\ \bibinfo {author} {\bibfnamefont
  {N.}~\bibnamefont {Yunes}},\ }\href {\doibase 10.1103/2ctd-mgb6} {\bibfield
  {journal} {\bibinfo  {journal} {Phys. Rev. D}\ }\textbf {\bibinfo {volume}
  {114}},\ \bibinfo {pages} {L021505} (\bibinfo {year} {2026})},\ \Eprint
  {http://arxiv.org/abs/2506.14695} {arXiv:2506.14695 [gr-qc]} \BibitemShut
  {NoStop}%
\bibitem [{\citenamefont {Carullo}\ \emph {et~al.}(2023)\citenamefont
  {Carullo}, \citenamefont {Del~Pozzo},\ and\ \citenamefont {Veitch}}]{pyRing}%
  \BibitemOpen
  \bibfield  {author} {\bibinfo {author} {\bibfnamefont {G.}~\bibnamefont
  {Carullo}}, \bibinfo {author} {\bibfnamefont {W.}~\bibnamefont {Del~Pozzo}},
  \ and\ \bibinfo {author} {\bibfnamefont {J.}~\bibnamefont {Veitch}},\ }\href
  {\doibase 10.5281/zenodo.8165507} {\enquote {\bibinfo {title}
  {\texttt{pyRing}: a time-domain ringdown analysis python package},}\
  }\bibinfo {howpublished}
  {\href{https://git.ligo.org/lscsoft/pyring}{git.ligo.org/lscsoft/pyring}}
  (\bibinfo {year} {2023})\BibitemShut {NoStop}%
\bibitem [{\citenamefont {Carullo}\ \emph {et~al.}(2019)\citenamefont
  {Carullo}, \citenamefont {Del~Pozzo},\ and\ \citenamefont
  {Veitch}}]{Carullo:2019flw}%
  \BibitemOpen
  \bibfield  {author} {\bibinfo {author} {\bibfnamefont {G.}~\bibnamefont
  {Carullo}}, \bibinfo {author} {\bibfnamefont {W.}~\bibnamefont {Del~Pozzo}},
  \ and\ \bibinfo {author} {\bibfnamefont {J.}~\bibnamefont {Veitch}},\ }\href
  {\doibase 10.1103/PhysRevD.99.123029} {\bibfield  {journal} {\bibinfo
  {journal} {Phys. Rev. D}\ }\textbf {\bibinfo {volume} {99}},\ \bibinfo
  {pages} {123029} (\bibinfo {year} {2019})},\ \bibinfo {note} {[Erratum:
  Phys.Rev.D 100, 089903 (2019)]},\ \Eprint {http://arxiv.org/abs/1902.07527}
  {arXiv:1902.07527 [gr-qc]} \BibitemShut {NoStop}%
\bibitem [{\citenamefont {Maenaut}\ \emph {et~al.}(2026)\citenamefont
  {Maenaut}, \citenamefont {Carullo}, \citenamefont {Cano}, \citenamefont
  {Liu}, \citenamefont {Cardoso}, \citenamefont {Hertog},\ and\ \citenamefont
  {Li}}]{Maenaut:2024oci}%
  \BibitemOpen
  \bibfield  {author} {\bibinfo {author} {\bibfnamefont {S.}~\bibnamefont
  {Maenaut}}, \bibinfo {author} {\bibfnamefont {G.}~\bibnamefont {Carullo}},
  \bibinfo {author} {\bibfnamefont {P.~A.}\ \bibnamefont {Cano}}, \bibinfo
  {author} {\bibfnamefont {A.}~\bibnamefont {Liu}}, \bibinfo {author}
  {\bibfnamefont {V.}~\bibnamefont {Cardoso}}, \bibinfo {author} {\bibfnamefont
  {T.}~\bibnamefont {Hertog}}, \ and\ \bibinfo {author} {\bibfnamefont
  {T.~G.~F.}\ \bibnamefont {Li}},\ }\href {\doibase 10.1103/k3hv-cqfn}
  {\bibfield  {journal} {\bibinfo  {journal} {Phys. Rev. D}\ }\textbf {\bibinfo
  {volume} {113}},\ \bibinfo {pages} {044039} (\bibinfo {year} {2026})},\
  \Eprint {http://arxiv.org/abs/2411.17893} {arXiv:2411.17893 [gr-qc]}
  \BibitemShut {NoStop}%
\bibitem [{\citenamefont {Abac}\ \emph {et~al.}(2026)\citenamefont {Abac} \emph
  {et~al.}}]{LIGOScientific:2025wao}%
  \BibitemOpen
  \bibfield  {author} {\bibinfo {author} {\bibfnamefont {A.~G.}\ \bibnamefont
  {Abac}} \emph {et~al.} (\bibinfo {collaboration} {LIGO Scientific, Virgo,
  KAGRA}),\ }\href {\doibase 10.1103/6c61-fm1n} {\bibfield  {journal} {\bibinfo
   {journal} {Phys. Rev. Lett.}\ }\textbf {\bibinfo {volume} {136}},\ \bibinfo
  {pages} {041403} (\bibinfo {year} {2026})},\ \Eprint
  {http://arxiv.org/abs/2509.08099} {arXiv:2509.08099 [gr-qc]} \BibitemShut
  {NoStop}%
\bibitem [{\citenamefont {Wagle}\ \emph {et~al.}(2026)\citenamefont {Wagle},
  \citenamefont {Li}, \citenamefont {Chen},\ and\ \citenamefont
  {Yunes}}]{Wagle:2026bpu}%
  \BibitemOpen
  \bibfield  {author} {\bibinfo {author} {\bibfnamefont {P.}~\bibnamefont
  {Wagle}}, \bibinfo {author} {\bibfnamefont {D.}~\bibnamefont {Li}}, \bibinfo
  {author} {\bibfnamefont {Y.}~\bibnamefont {Chen}}, \ and\ \bibinfo {author}
  {\bibfnamefont {N.}~\bibnamefont {Yunes}},\ }\href@noop {} {\  (\bibinfo
  {year} {2026})},\ \Eprint {http://arxiv.org/abs/2610.27116} {arXiv:2610.27116
  [gr-qc]} \BibitemShut {NoStop}%
\end{thebibliography}%

\clearpage
\onecolumngrid
\newpage
\begin{center}
  \textbf{\large{Supplemental Material}} \\
\end{center}

\twocolumngrid

\section{Validity guards, posterior reweighting, and combination across events}
\label{sec:guard_reweighting}

The quasinormal-mode fits and perturbative expansion used in this analysis are restricted to a finite region of parameter space. We enforce these restrictions during the likelihood evaluation through a validity guard. 
As some of the guard conditions depend on event-specific source parameters, the fraction of the nuisance-prior volume satisfying the validity conditions depends on $(\ell,\mu)$, which are common parameters across events. 
The posterior returned directly by the sampler therefore contains a parameter-dependent prior-volume contribution that must be removed before the posteriors on $\ell$ and $\mu$ are combined.

\subsection*{Posterior reweighting}

For event $i$, we denote the common theory parameters by
\begin{equation}
    \lambda=(\ell,\mu),
\end{equation}
and the event-specific parameters by $\boldsymbol{\eta}_i$. 
The likelihood sampled in the production analysis is effectively
\begin{equation}
    {\cal L}_i(d_i|\lambda,\boldsymbol{\eta}_i)
    \longrightarrow
    {\cal L}_i(d_i|\lambda,\boldsymbol{\eta}_i)
    G_i(\lambda,\boldsymbol{\eta}_i),
    \label{eq:SM_guard_likelihood}
\end{equation}
where $G_i=1$ when all validity conditions are satisfied and $G_i=0$ otherwise.

Schematically, the guard takes the form
\begin{align}
    G_i(\ell,\mu,\boldsymbol{\eta}_i)
    ={}&
    \Theta\!\left(M_{{\rm src},i}-\ell\right)
    \Theta\!\left(1-M_{{\rm src},i}\mu\right)
    \nonumber\\
    &\times
    \prod_{q\in{\cal Q}_i}
    \Theta\!\left[-{\rm Im}\,\omega^{(A)}_{q}\right]
    \Theta\!\left[-{\rm Im}\,\omega^{(P)}_{q}\right],
    \label{eq:SM_massive_guard}
\end{align}
where ${\cal Q}_i$ denotes the set of quasinormal modes included in the analysis of event $i$. 
The first two conditions impose the perturbative and massive-scalar fitting domains, whereas the remaining conditions require the relevant axial and polar modes to remain decaying.

At fixed $(\ell,\mu)$, the fraction of the event-specific nuisance prior satisfying the guard is
\begin{equation}
    N_i(\ell,\mu)
    =
    \int d\boldsymbol{\eta}_i\,
    \pi_i(\boldsymbol{\eta}_i|\ell,\mu)\,
    G_i(\ell,\mu,\boldsymbol{\eta}_i).
    \label{eq:SM_guard_fraction}
\end{equation}
Consequently, even though the intended prior $\pi(\ell,\mu)$ is uniform, imposing the guard gives an effective marginal prior proportional to
\begin{equation}
    \pi(\ell,\mu)\,N_i(\ell,\mu).
\end{equation}

It is useful to define the normalized nuisance prior conditioned on validity,
\begin{equation}
    \pi_i^{G}(\boldsymbol{\eta}_i|\ell,\mu)
    =
    \frac{
        \pi_i(\boldsymbol{\eta}_i|\ell,\mu)
        G_i(\ell,\mu,\boldsymbol{\eta}_i)
    }{
        N_i(\ell,\mu)
    }.
    \label{eq:SM_normalized_guard_prior}
\end{equation}
The corresponding event-level likelihood marginalized over the validity-conditioned nuisance prior is
\begin{equation}
\begin{split}
& \hat{{\cal L}}_i(\ell,\mu)\\
& =
    \frac{1}{N_i(\ell,\mu)}
    \int d\boldsymbol{\eta}_i\,
    {\cal L}_i(d_i|\ell,\mu,\boldsymbol{\eta}_i)
    \pi_i(\boldsymbol{\eta}_i|\ell,\mu)
    G_i(\ell,\mu,\boldsymbol{\eta}_i).
\end{split}
\label{eq:SM_corrected_likelihood}
\end{equation}
The posterior returned directly by the production sampler therefore satisfies
\begin{equation}
    \tilde p_i(\ell,\mu|d_i)
    \propto
    \pi(\ell,\mu)\,
    N_i(\ell,\mu)\,
   \hat{{\cal L}}_i(\ell,\mu),
\end{equation}
whereas the posterior associated with the normalized conditional prior is
\begin{equation}
    \hat p_i(\ell,\mu|d_i)
    \propto
    \frac{
        \tilde p_i(\ell,\mu|d_i)
    }{
        N_i(\ell,\mu)
    }.
    \label{eq:SM_guard_reweighting}
\end{equation}

\subsection*{Numerical evaluation of the guard fraction}

We evaluate $N_i(\ell,\mu)$ numerically by Monte Carlo integration over the same event-specific nuisance prior used in the production analysis. For each event, we generate a fixed set of $N_{\rm MC}=200000$ nuisance-prior samples,
\begin{equation}
    \left\{
    \boldsymbol{\eta}_{i,j}
    \right\}_{j=1}^{N_{\rm MC}}
    \sim
    \pi_i(\boldsymbol{\eta}_i|\ell,\mu).
\end{equation}
Only nuisance parameters entering the guard need to be sampled explicitly; integrations over nuisance parameters on which $G_i$ does not depend reduce to unity.

For each point $(\ell_a,\mu_b)$ on a rectangular grid covering the theory-prior support, we evaluate the complete production guard for every nuisance-prior realization. 
The Monte Carlo estimate of the guard fraction is then
\begin{equation}
    \widehat N_i(\ell_a,\mu_b)
    =
    \frac{1}{N_{\rm MC}}
    \sum_{j=1}^{N_{\rm MC}}
    G_i(\ell_a,\mu_b,\boldsymbol{\eta}_{i,j}).
    \label{eq:SM_guard_MC}
\end{equation}
Thus, $\widehat N_i$ is simply the fraction of nuisance-prior realizations that remain valid at the specified theory parameters. If the nuisance samples are generated from an importance distribution $q_i$ rather than directly from $\pi_i$, we instead use
\begin{equation}
    \widehat N_i(\ell_a,\mu_b)
    =
    \frac{
        \sum_j r_{i,j}
        G_i(\ell_a,\mu_b,\boldsymbol{\eta}_{i,j})
    }{
        \sum_j r_{i,j}
    },
\end{equation}
and 
\begin{equation}
    r_{i,j}
    =
    \frac{
        \pi_i(\boldsymbol{\eta}_{i,j}|\ell_a,\mu_b)
    }{
        q_i(\boldsymbol{\eta}_{i,j})
    }.
\end{equation}

We evaluate $\widehat N_i$ on a grid of
$301\times181$ points in $(\ell,\mu)$ and interpolate the resulting two-dimensional guard-fraction surface to the locations of the nested-sampling posterior samples. 
The same source-frame conversion, quasinormal-mode fits, and validity conditions used in the production likelihood are used when evaluating the guard. 
Using the same nuisance-prior realization at every $(\ell,\mu)$ grid point also avoids introducing independent Monte Carlo fluctuations between neighboring grid cells.

For a production posterior sample $n$ at $(\ell_n,\mu_n)$ with original nested-sampling weight $\tilde w_{i,n}$, we assign the reweighted value
\begin{equation}
    w_{i,n}^{\rm rw}
    =
    \frac{
        \tilde w_{i,n}/
        \widehat N_i(\ell_n,\mu_n)
    }{
        \displaystyle
        \sum_m
        \tilde w_{i,m}/
        \widehat N_i(\ell_m,\mu_m)
    }.
    \label{eq:SM_sample_reweight}
\end{equation}
These normalized weights are then used to construct the one- and two-dimensional reweighted posterior distributions shown in this work.

The reweighting is defined only where
$\widehat N_i(\ell,\mu)>0$.
Regions for which $\widehat N_i=0$ have no support under the validity-conditioned model and are not restored by the reweighting procedure. A boundary caused by $\widehat N_i=0$ therefore reflects the adopted validity domain rather than information provided by the gravitational-wave data.

\subsection*{Combination across events}

Since $\ell$ and $\mu$ characterize the underlying theory, we assume that they take the same values for all observed black-hole remnants. For statistically independent events,
\begin{equation}
    p(\ell,\mu|\{d_i\})
    \propto
    \pi(\ell,\mu)
    \prod_{i=1}^{N_{\rm ev}}
   \hat{{\cal L}}_i(\ell,\mu).
    \label{eq:SM_combined_posterior}
\end{equation}
Using the reweighted event-level posteriors,
\begin{equation}
    \hat p_i(\ell,\mu|d_i)
    \propto
    \pi(\ell,\mu)
   \hat{{\cal L}}_i(\ell,\mu),
\end{equation}
the combined posterior becomes
\begin{equation}
    p(\ell,\mu|\{d_i\})
    \propto
    \pi(\ell,\mu)^{1-N_{\rm ev}}
    \prod_{i=1}^{N_{\rm ev}}
    \hat p_i(\ell,\mu|d_i).
    \label{eq:SM_combined_from_posteriors}
\end{equation}
For the common uniform prior on $(\ell,\mu)$ adopted here, the prior factor is constant within its support and the combined posterior is therefore proportional to the product of the reweighted event-level posteriors.

\section{Event-by-event Results}

\subsection{Posterior distribution of $\zeta$}

\begin{figure*}[t]
    \centering
    \includegraphics[width=\textwidth]
    {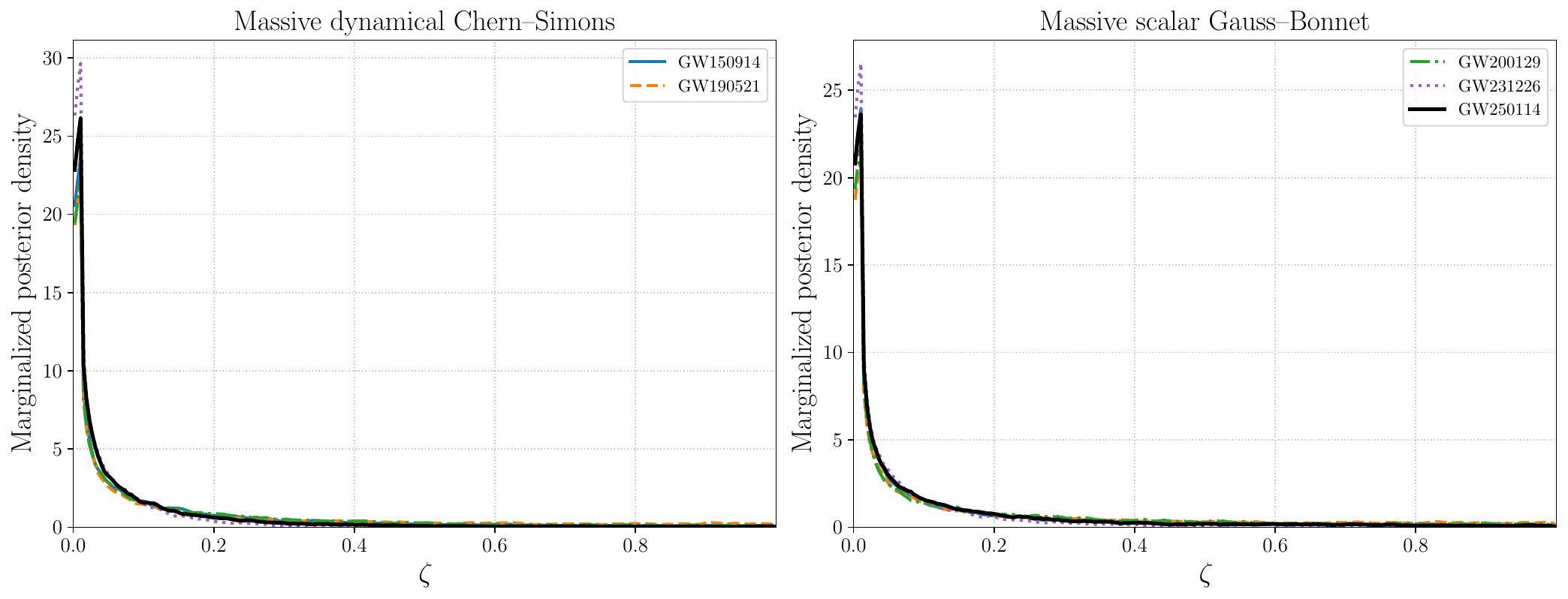}
    \caption{
    Marginalized posterior distributions of $\zeta$ for massive scalar fields with dynamical Chern--Simons (left) and scalar Gauss--Bonnet (right) couplings, obtained from Bayesian inference of the individual ringdown events analyzed in this work.
    }
    \label{fig:zeta_posteriors}
\end{figure*}

Figure~\ref{fig:zeta_posteriors} shows the marginalized posterior distributions of $\zeta$ [defined in Eq.~(\ref{eq:zeta})] for massive scalar fields with dynamical Chern--Simons (left) and scalar Gauss--Bonnet (right) couplings, obtained from the individual ringdown signals. 
The posterior support is concentrated predominantly at $\zeta\lesssim0.2$, well below the validity boundary $\ell=M$, corresponding to $\zeta=1$. 
This indicates that the constraints on $\ell$ are not driven by truncation at the validity boundary. 
The range of $\zeta$ favored by the posteriors is also consistent with the forecast reported in~\cite{Lui:2026osx}.

\subsection{Reweighted posterior}

\begin{figure*}[t]
    \centering
    \includegraphics[width=\textwidth]
    {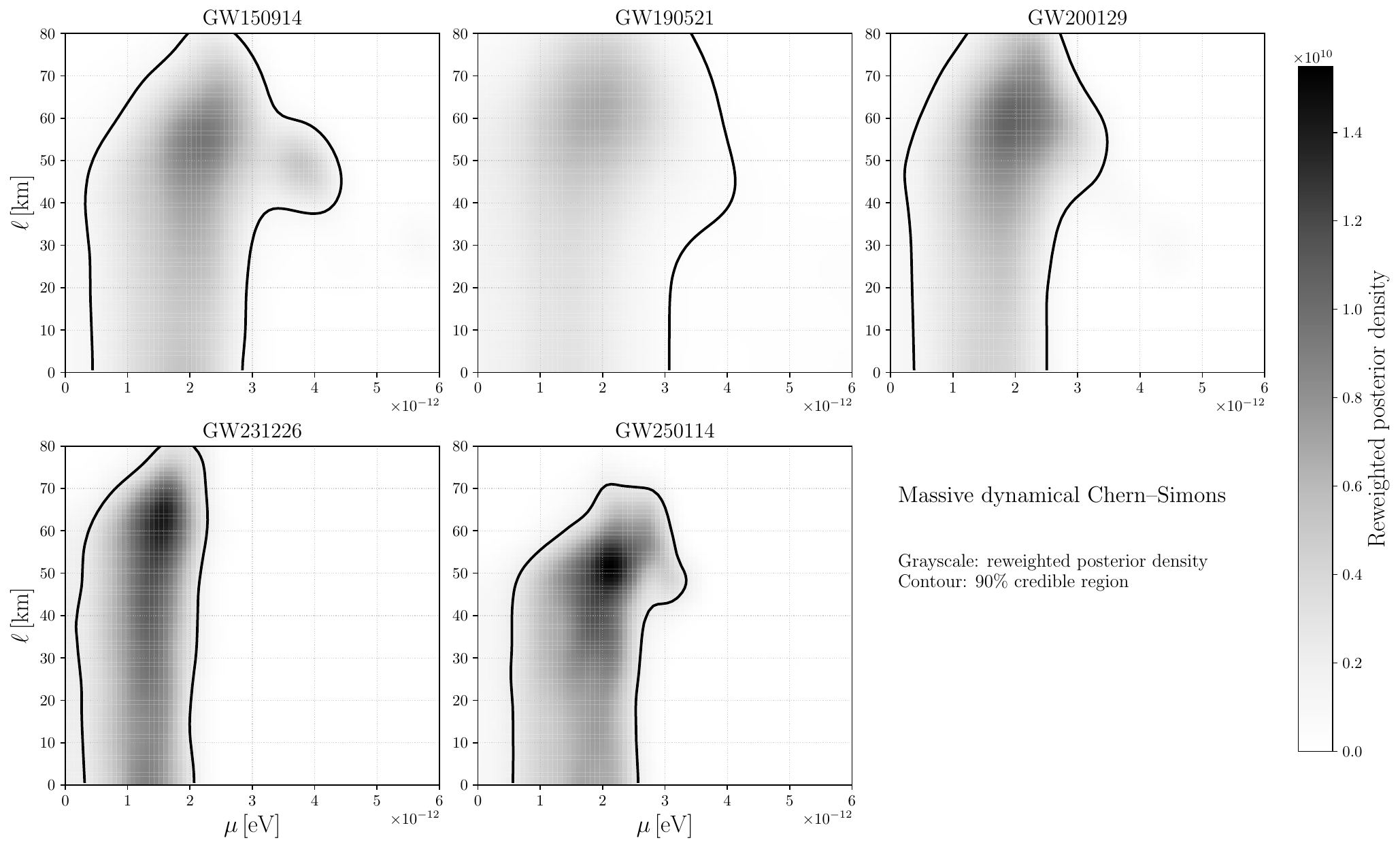}
    \includegraphics[width=\textwidth]
    {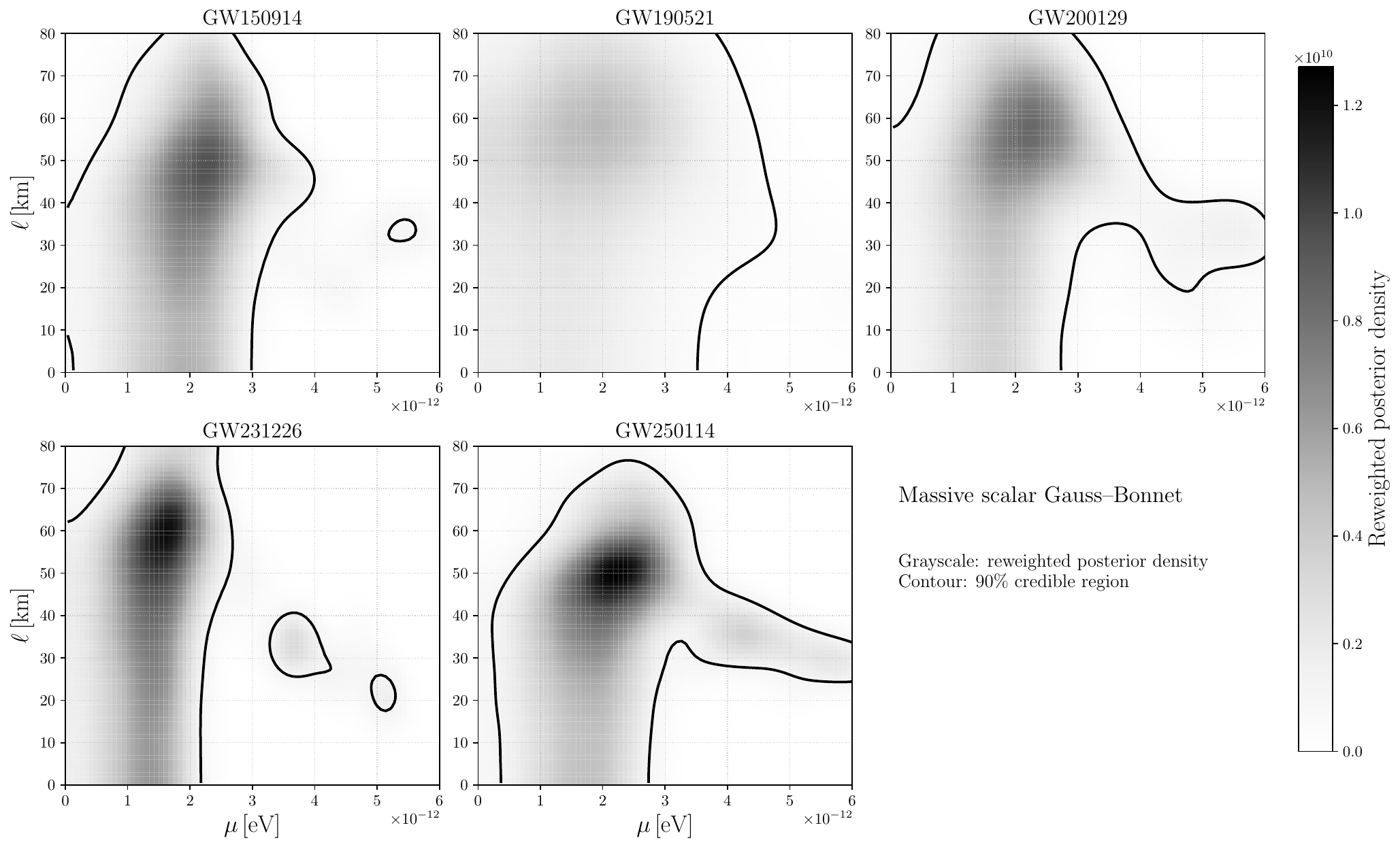}
    \caption{
    Event-by-event reweighted posterior distributions in the $(\ell,\mu)$ plane for massive scalar fields with dynamical Chern--Simons (top) and scalar Gauss--Bonnet (bottom) couplings.
    The grayscale shows the reweighted posterior density, while the contour marks the $90\%$ credible region for each event.
    }
    \label{fig:SM_event_posteriors}
\end{figure*}

Figure~\ref{fig:SM_event_posteriors} shows the event-by-event reweighted posterior distributions in the $(\ell,\mu)$ plane for dynamical Chern--Simons and scalar Gauss--Bonnet couplings, respectively. 
The contours indicate the corresponding $90\%$ credible regions.

\setcounter{equation}{0}
\setcounter{figure}{0}
\setcounter{table}{0}
\setcounter{page}{1}
\makeatletter
\renewcommand{\theequation}{S\arabic{equation}}
\renewcommand{\thefigure}{S\arabic{figure}}
\renewcommand{\thetable}{S\arabic{table}}



\end{document}